\documentclass [12pt]{article}
\usepackage{graphicx,amssymb,amsfonts,latexsym,amsmath,amsthm,times}
\usepackage{epsfig}
\usepackage{fancyhdr}
\usepackage{color}
\usepackage[english]{babel}
\newcommand{\mfrac}[2]{\displaystyle \frac{#1}{#2}}

\numberwithin{equation}{section}

\begin{document}

\setcounter{page}{42}

\footnotesize {\flushleft \mbox{\bf \textit{Math. Model. Nat. Phenom.}}}
 \\
\mbox{\textit{{\bf Vol.~3, No.~3, 2008, pp.~42-86}}}

\thispagestyle{plain}

\vspace*{2cm} \normalsize \centerline{\Large \bf Population growth and persistence in a heterogeneous}
\centerline{\Large \bf environment: the role of diffusion and advection}

\vspace*{1cm}

\centerline{\bf A. B. Ryabov $^a$\footnote{Corresponding author. E-mail: a.ryabov@icbm.de} and B. Blasius $^a$}

\vspace*{0.5cm}

\centerline{$^a$ Institute for Chemistry and Biology of Marine Environment (ICBM),}
\centerline{Oldenburg University, 26111 Oldenburg, Germany}

\vspace*{1cm}

\noindent {\bf Abstract.}
The spatio-temporal dynamics of a population present one of the most fascinating aspects and challenges for ecological modelling. In this article we review some simple mathematical models, based on one dimensional reaction-diffusion-advection equations, for the growth of a population on a heterogeneous habitat. Considering a number of models of increasing complexity we investigate the often contrary roles of advection and diffusion for the persistence of the population. When it is possible we demonstrate basic mathematical techniques and give the critical conditions providing the survival of a population in simple systems and in more complex resource-consumer models which describe the dynamics of phytoplankton in a water column.

\vspace*{0.5cm}

\noindent {\bf Key words:} diffusion, advection, survival, population, phytoplankton.

\noindent {\bf AMS subject classification:} 35K57, 76R10, 76R50, 92D25, 92D40

\vspace*{1cm}

\setcounter{equation}{0}

\section{Introduction}

For long times field biologists and naturalists have been fascinated by the richness and beauty of complex patterns that can be observed in
spatially extended populations. However, the same observations also constitute a challenge for theoreticians who aim to explain this complexity by means of
mathematical models. At first glance one might be tempted to argue that
the spatial diversity of natural populations mainly originates from some underlying abiotic
heterogeneity of the environment. If growth conditions vary between different locations then this spatial variation should be reflected in the density distribution of natural populations. Thus, it is reasonable to assume that a large part of the observed richness in the patterning of biotic landscape can be attributed to the spatial heterogeneity of growth conditions.
On the other hand in nearly all systems populations at different locations are coupled and are able to interact with each other, be it by random, diffusive mixing or by directed, advective flows or currents, which are able to transport individual organisms from one region to another.
Note that the term ``advection'' here is used in a very general sense, including different mechanisms of transport, such as drift caused by a current of water, sinking and floating-up in the gravity field, chemotaxis, etc.

The impact of such spatial interactions on the fate of a local population can be very diverse, depending on whether the flow brings-in new immigrants into the habitat or if it takes them away.
Furthermore, diffusion and advection can have opposing influences, where, for example, one process may be beneficial for the population, while another process has adverse effects.
Usually, the role of an advective flux is evident and mainly depends on its strength and direction. In contrast, the role of diffusivity can be of two kinds. On the one hand, diffusivity accelerates the spread of a population in a habitat and is necessary to provide the population's persistence in a flux. On the other hand, intensive mixing may transport too many organisms into unfavourable zones, resulting in the extinction of the population.
Moreover, in consumer-resource models usually also the resource fluxes are driven by diffusivity and advection, a fact which leads to even new patterns and dynamical behaviour.
As a consequence, it is quite possible to find stable populations in locations where growth conditions alone would not permit persistence. On the other hand, seemingly well-being habitats may not be able to sustain a stable population, if they suffer population outflows.
From all these effects, populations are spatially structured in an intricate interplay between local growth and its geographical variations on the one hand and spatial transport by diffusion and advection on the other hand.

This article is devoted to present an introductory review of these topics and to introduce the reader
some of the most commonly used models for the population dynamics in a non-uniform turbulent environment.
Even though we have in particular the specific case of phytoplankton dynamics in mind,
the main concepts considered here hold for any population subjected to mixing and advection.
When it is possible, to reproduce the whole picture we perform derivations of the main mathematical relations. Otherwise we aim at least to demonstrate some basic ideas and refer to detailed discussions, providing information about the main techniques that are useful in this field of research.
There exist many excellent reviews about spatial population dynamics (see e.g. \cite{holmes_partial_1994, klausmeier_spatial_2002, neuhauser_mathematical_2001, okubo_diffusion_2001}).
Nevertheless to our knowledge, the interplay of advection and diffusion for a heterogeneous population, as elaborated in this text, has never been described.
So, while the text will not provide much new insights for the specialist, we hope that many readers will find this text a valuable introduction into the basic mechanism and critical conditions providing the survival of a population in a heterogeneous environment.

To focus on the main ideas, we had to confine the review to topics related with population survival. However, we would like to briefly mention other important aspects of spatial population dynamics
that had to be omitted here. First, we had to restrict to single species populations, while
one of the most important and still debatable challenges concerns the competition of spatially structured populations
\cite{tilman_resource_1982, grover_resource_1990, sommer_competition_2002} and the high diversity of phytoplankton species (see e.g. \cite{roy_s._towards_2007} and the references therein). Second, for the sake of simplicity we included only Fickian diffusion models, whereas in the ocean diffusivity depends on the scale of phenomena \cite{okubo_diffusion_2001}.
Third, we had to omit the interesting aspects concerning conditional persistence in situations where the growth is influenced by an Allee effect \cite{stephens_consequences_1999}. Fourth, the interaction of turbulent currents  with other processes can produce complex spatial structures in phytoplankton distributions \cite{martin_phytoplankton_2003}, whereas locally oscillatory behaviour may lead to spatio-temporal chaotic waves \cite{petrovskii_minimal_1999, petrovskii_wave_2001}. Finally, we did not even touch such important problems as model validation and the simulation of multicomponent natural ecosystems \cite{moll_review_2003}.

The review is structured as follows.
In chapter two, we shortly discuss some non-spatial models, which will be used later to build-up spatial explicit models.
In chapter three we consider critical patch models, starting from the well known KiSS model, and proceed with more sophisticate models, which draw out the influence of boundary conditions, finiteness of the favourable patch, heterogeneity of the mixing, etc. Furthermore, we consider the Fisher-Kolmogorov equation and its extensions to advection, non-uniform mixing and heterogeneous environments.
In the last chapter, we consider 1D models of the vertical phytoplankton distribution, starting from those that include only light limitation and then considering models including many limiting factors. However, even in the latter models we focus on the basic theoretical aspects and general conclusions.

\setcounter{equation}{0}
\section{Non-spatial models \label{Sec:non_spatial}}

\paragraph{Logistic growth}

As a warming-up, in this chapter we shortly describe some simple population models in a non-spatial context.
These models will mainly be used as a base for developing spatially extended models,
but they may also serve as a benchmark for exploring the properties of their spatial analogues.
Let $P(t)$ denote the density of a population of interest at time $t$.
In the simplest way, the dynamics of $P$ can be modelled in terms of an ordinary differential equation of first order
\begin{eqnarray}
\frac{dP}{dt} = \mu (P) P \ ,
\label{eq:genGrowth}
\end{eqnarray}
where the growth rate $\mu(P)$ depends on the population density $P$. The function $\mu(P)$ takes into account for density dependent regulatory factors which
usually are associated with resource depletion or conditions of over-crowding.
Consequently, (neglecting the possibility of an Allee effect) we may assume that  $\mu(P)$
is monotonically decreasing and eventually becomes negative for large densities.
The simplest form to put this into a model is the logistic growth
\begin{eqnarray}
\frac{dP}{dt} = \mu_0 P \left( 1 - \frac{P}{K}\right) ,
\label{eq:logGrowth}
\end{eqnarray}
where $K$ is the carrying capacity of the system and describes the maximal population density that can be sustained by the system, i.e. $\mu(K)=0$.
Equation (\ref{eq:logGrowth}) can be thought to arise from a Taylor expansion of equation \eqref{eq:genGrowth} for small densities $P\ll K$, where $\mu_0\equiv\mu(0)$ is the growth rate at small density and $K \equiv -\mu_0 \left(  \partial\mu / \partial P  \right)^{-1} $ is the carrying capacity.

\paragraph{Resource-limited growth}
As a secondary model class, we investigate resource limited population growth,
where the dynamics of the resource is explicitly included into the model equations.
Possibly the most simple way to model an isolated consumer-resource system (i.e., in the absence of any fluxes out of the system) is given in the following form
\begin{eqnarray}
\frac{dP}{dt} & = & \quad \mu (N) P -m P \ ,\nonumber \\ 
\label{eq:NP}\\
\frac{dN}{dt} & = & - \mu (N) P +m P \nonumber \ .
\end{eqnarray}
Here $N(t)$ denotes the resource concentration (e.g., a limiting nutrient such as nitrate or phosphate) at time $t$ and the population abundance $P(t)$ is measured in units of the organism's resource content. As usual, in contrast to equation (\ref{eq:genGrowth}), here we explicitly separated the growth into a term proportional to the resource uptake, $\mu(N) P$, and a death process with mortality $m$.
Note that in this review the symbol $\mu$ has a somewhat different interpretation in the context of a resource-implicit (\ref{eq:genGrowth}) or resource-explicit model (\ref{eq:NP}). We hope that the reader is not confused by this notation.

\begin{figure}[tb]
\centerline{\includegraphics[height=5cm]{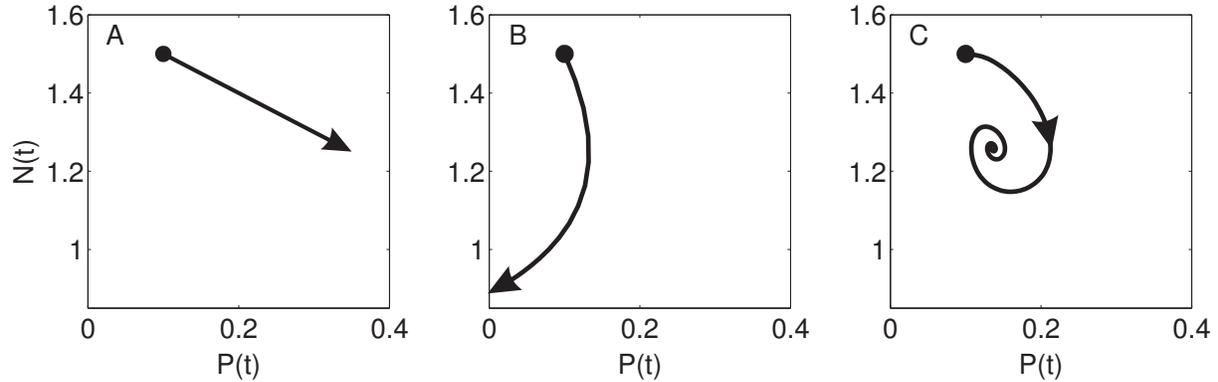}
}
\caption{Typical phase trajectories of the non-spatial models. (A) the dynamics
of the conservative
model \eqref{eq:NP} is reduced to a straight line; (B) the non-conservative
model \eqref{eq:NP_noncons} leads to the
extinction of the population; (C) the trajectory of system \eqref{Eq:EffOneDimChem}
spirals in phase space to an isolated fixed point. The initial values
P(0) = 0.1, N(0) = 1.5 are marked by filled circles, other
parameters are $c=5$, $H_N=5$, $m = 1$, $\delta=0.005$, $\varepsilon = 0.5$, and $N_i = 15$.} \label{fig:mixed_mod}
\end{figure}

Usually, $\mu(N)$ is assumed
to vanish for small resource concentrations ($\mu(0)=0$) and to be saturated for large $N$. A convenient parametrisation is given by the Monod form
\begin{equation}
\mu(N)= c \, \frac{N}{H_N+N}
\end{equation}
with the half-saturation constant $H_N$.

Equation (\ref{eq:NP}) represents a closed system, where recycling is assumed to be 100\% efficient. Thus the biomass that is taken out of the system due to death processes is remineralised into the nutrient pool, i.e. no material is lost in the conversion from consumer to nutrients. As a consequence, this system can be reduced to a one-dimensional model, despite the fact that it contains two equations.
The reason is that the total resource concentration,
either free or bound to the population, is a first integral of motion: $S=N + P=\mathrm{const}$.
After plugging this into the equation for $P$ we obtain a single equation of the form (\ref{eq:genGrowth}) for the time dynamics of the population density,
\begin{equation}
\frac{dP}{dt}=P\, [\mu(S-P) -m] \ .
\label{Eq:EffOneDim}
\end{equation}
If the saturation of the uptake rate is neglected, $\mu(N)=c N$, we retrieve the logistic growth model (\ref{eq:logGrowth}) with
\begin{equation}
\mu_0=cS-m \quad \mathrm{and} \qquad K=S-\frac{m}{c}.
\end{equation}
The phase trajectory of system \eqref{eq:NP} in $(P, N)$ coordinate is a straight line from the initial conditions to an equilibrium state (Fig.~\ref{fig:mixed_mod}A).

We note though, that the reduction to a single equation works only for a closed system. In general, recycling efficiency will never be fully effective.
Thus, one may assume that only a fraction $\varepsilon$ of the dead biomass is remineralised, leading to a non-conservative model
\begin{eqnarray}
\frac{dP}{dt} & = & \quad \mu (N) P -m P \ ,\nonumber  \\[-3mm]
\label{eq:NP_noncons} \\
\frac{dN}{dt} & = & - \mu (N) P + \varepsilon m P \nonumber \ .
\end{eqnarray}
The two approaches introduced in this section will be used in chapter 4 as building blocks to model the local reaction of a spatial extended population. However we should note that both  (\ref{eq:NP}) and   (\ref{eq:NP_noncons})  have some shortcomings, which ultimately are related to the fact that the models do not contain in- and out-flows. For example, the model (\ref{eq:NP_noncons}) does not even admit isolated stationary states in the $(N,P)$-phase plane and for any initial condition leads to the extinction of the population, $P(t) \to 0$ (Fig.~\ref{fig:mixed_mod}B). While this is no problem in the spatial extended counterparts of the models, which necessarily have exchanges with surroundings specified by the boundary conditions, in local models these exchanges should be introduced as additional terms.

\paragraph{Chemostat models}

To show that these problem disappear as soon as external fluxes are incorporated into the models, we shortly introduce the most simple and frequently used chemostat model.
Consider a well-stirred reactor that contains phytoplankton cells with concentration $P(t)$ and a limiting nutrient with concentration $N(t)$. The chemostat is supplied with the nutrient at an input concentration $N_i$ from an external nutritive medium. The outflow contains both medium and phytoplankton cells. Inflow and outflow
are characterised by the dilution rate $\delta$. Then the chemostat equations take the following form
\begin{eqnarray}
\frac{dP}{dt} & = &\mu (N) P -(m+ \delta) P \ ,\nonumber \\[-3mm]
\label{eq:Chemost} \\
\frac{dN}{dt} & = & \delta(N_i -N) - \mu (N) P + \varepsilon m P \nonumber\ .
\end{eqnarray}
In this model, the total amount of free and bound nutrients $S=N+P$ is  not conserved.
Thus, the chemostat sustains
a stable feasible equilibrium and the system spirals in phase space to an isolated fixed point (Fig.~\ref{fig:mixed_mod}C), see e.g., \cite{huppert__2002, clodong_chaos_2004, huppert_what_2004}.
Furthermore, for perfect nutrient recycling ($\varepsilon=1$) one can easily show that $S(t)$ follows the simple dynamics
\begin{equation}
\frac{dS}{dt}=\delta (N_i-S) \ .
\end{equation}
Therefore, after some transient time, determined by the exchange rate $\delta$, the system settles to a state where $S=N_i$. This means that the total amount of nutrients in the system, either bound to the population or free, asymptotically goes to the external nutrient concentration, and
one asymptotically retains a one-dimensional equation, similar to (\ref{Eq:EffOneDim})
\begin{equation}
\frac{dP}{dt}=
P\, [ \mu (N_i-P)  -(m+ \delta)]  \ .
\label{Eq:EffOneDimChem}
\end{equation}

\setcounter{equation}{0}

\section{Critical patch models for an extended population}

In this section we review some simple one-dimensional spatio-temporal models which demonstrate the intricate interplay between mixing, advection, boundary conditions and other factors which are critical for the persistence of a population.
Consider a population growing on a one-dimensional (1D) axis, which may represent either a horizontal or vertical spatial extension.
Let $P(x, t)$ denote the population density at time $t$ and position $x$.  We assume that the change of density occurs as a result of the local death-birth processes and of the spatial movement of organisms due to diffusion and advection.
The dynamics of such a population  can be written in terms of a reaction-diffusion-advection equation \cite{okubo_diffusion_2001, murray_mathematical_2003}
\begin{eqnarray}
\frac{\partial P(x, t) }{\partial t} &=&  \mbox{reproduction} - \mbox{advection} + \mbox{mixing}
\nonumber \\
\nonumber \\
&=& \mu(x, P)  P -  v \frac{ \partial P }{\partial x} + \frac{\partial  }{\partial x} D \frac{\partial P }{\partial x}
\label{eq:rdiff}\ ,
\end{eqnarray}
where $\mu$ represents the growth rate (compare to equation (\ref{eq:genGrowth})), $v$ is the advection velocity
and $D$ is the diffusivity, which can depend on $x$ and other variables.
The advective term $v \frac{ \partial P }{\partial x} $ describes the drift of the population with the constant velocity $v$. 
There are many physical or biological factors which can give rise to advective processes.
For instance, in a vertical system the organisms (e.g. phytoplankton cells) may sink (or rise) because they are more heavy (or light) than their surrounding medium. Another example arises for a population in  a vertical/horizontal stream (e.g., in a horizontal flow of a river or in an ocean current).
Advection and diffusion can be written in terms of a derivative $\frac{\partial}{\partial x} J(P,x)$ so that the whole flux of biomass is given as
\begin{equation}
J = v P - D\frac{\partial P }{\partial x} \ .
\end{equation}

Additionally, model (\ref{eq:rdiff}) has to be complemented by appropriate boundary conditions which specify the environmental influence on the spatially extended system.
One can distinguish between three important types of boundary conditions.
The first type (or Dirichlet) boundary condition merely specifies the value of the solution at the boundary. For example, the condition
\begin{eqnarray}
P(0) = 0,\  P(L) = 0
\end{eqnarray}
assumes that the population density vanishes at the habitat edges, which corresponds to an absolutely hostile environment outside the segment $[0, L]$.
The second type (or Neumann) boundary condition specifies the flux across the boundary. For instance, the condition
\begin{eqnarray}
J(x)|_{x = 0, L} = \left( \left.v P - D \frac{\partial P }{\partial x} \right)  \right|_{x = 0, L} = {\rm 0}
\end{eqnarray}
specifies impenetrable boundaries, in other words there is no in- and outflow of biomass across the habitat edges.
The boundary condition of the third (Robin) type is a linear combination of the Dirichlet and Neumann conditions. Consider, for instance, a penetrable barrier, where
the flux of  organisms across the barrier is proportional to the difference of the population density outside, $P_{out}$, and inside, $P(x)$, yielding
\begin{eqnarray}
\left. J(x) \right|_{x = 0, L} =\left( \left.v P - D \frac{\partial P }{\partial x} \right)  \right|_{x = 0, L}= \left. h (P(x) - P_{out}) \right|_{x = 0, L}\ .
\end{eqnarray}
If the permeability of the barrier $h = 0$ this condition is equivalent to the zero flux boundary condition, and as $h\rightarrow \infty$ it approaches the Dirichlet boundary conditions $P(L) = P_{out}$.

\subsection{Persistence on a finite patch: critical patch size and mixing \label{sec:crit}}

For simplicity, we begin our investigation with models which implement only diffusion but no advection.
We concentrate on the population dynamics on a finite favourable patch and
show that in this case diffusion plays a negative role for the populations' persistence.

In the absence of advection, $v=0$, equation (\ref{eq:rdiff}) reduces to
\begin{equation}
\frac{\partial P(x, t) }{\partial t} = \mu(x, P)  P  +
D \frac{\partial^2 P}{\partial x^2}
\label{eq:rdiffNoAd}\
\end{equation}
complemented by appropriate boundary conditions.
Perhaps the most important biological question that one may ask in this model concerns the conditions under which the population will be able to persist or die out.
This question can be trivially answered for an infinite homogeneous environment, where the growth rate is independent of the spatial coordinate, $\mu(x, P)=\mu(P)$.
In this case (and assuming that there is no multistability, i.e.
the equation $\mu(P) = 0$ has a unique positive solution $P^*$)
the fate of the population entirely depends on the sign of the linearisation of the growth rate around $P=0$.
The population can invade if and only if the linearised growth rate $\mu_0 = \mu(P = 0) > 0$. Furthermore, it is easy to verify that the final distribution must be uniform, $P(x) = P^*$.

In reality the growth conditions will differ between various locations so that $\mu(x, P)$ will depend on $x$. We refer to such a situation as a ``heterogeneous environment''.
In a heterogeneous system, obviously also the linearised growth rate will be a function of the spatial position, $\mu_0 = \mu_0(x)$, which makes the problem of the population persistence more complicated.
Even if locally the population faces bad environmental conditions, $\mu_0(x)<0$, due to the population inflow from other locations it may still be able to persist. On the other hand, a favourable patch with $\mu_0(x)>0$ is no longer save, because diffusion out off the patch can cause additional losses which may even lead to local extinction.

\begin{figure}[tb]
\centerline{\includegraphics[height=5cm]{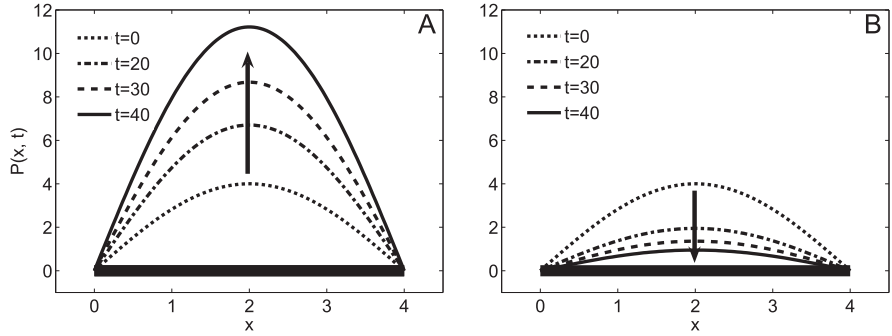}
}
\caption{Influence of diffusion on the persistence of an isolated population in the KiSS
model (\ref{eq:KiSSeq}). (A) $D = 3.2 < D_0$, the population growth is unbounded; (B) $D = 3.3 > D_0$, the population goes extinct.
Parameters are $\mu_0=2$, $L=4$, $K = 400$, thus $D_0 = 3.24$. The horizontal thick black line shows the location of the favourable patch. } \label{fig:skel}
\end{figure}

\paragraph{KiSS model}
One simple and elegant model to study this problem
has been independently introduced by Kierstead and Slobodkin \cite{kierstead_size_1953} and Skellam \cite{skellam_random_1951}. The main idea is to separate the landscape into a favourable area (which will be denoted as the species' habitat) adjoining some hostile environment from both sides.
For its simplicity, Akiro Okubo assigned to the model the name KiSS, which on the one hand includes the authors' initials, and from the other hand can be deciphered as ``Keep it Simple Stupid'' \cite{slobodkin_akira_1999}.
The KiSS model suggests a population growing on a finite patch of size $L$, surrounded by an absolutely hostile environment with infinite mortality (Fig.~\ref{fig:skel}).
Thus,  per definition the population density outside of the favourable patch equals zero, and it is sufficient to consider the dynamics inside the segment $0 \leq x \leq L$, upon the condition that the solution vanishes at the boundaries. To simplify the situation even further, the growth term
in equation (\ref{eq:rdiffNoAd}) is linearised for small density and is set constant $\mu_0$ within the habitat, which yields the following equations
\begin{eqnarray}
\frac{\partial P }{\partial t}
&=&  \mu_0 P +  D \frac{\partial^2 P}{\partial x^2} \ , \qquad 0 \leq x \leq L
\label{eq:KiSSeq}\\
  \nonumber\\
P(0) &=& P(L)  = 0 \ . \label{eq:KiSSbc}
\end{eqnarray}
This model could represent, for instance, the vegetation patterns of coastal plants growing on some island in the ocean, where diffusion of plants takes place via seed dispersal. Seeds which are transported out of the island into the water cannot survive, which is equivalent to infinite mortality outside of the favourable patch.

A detailed analysis of the KiSS model can be found in  \cite{skellam_random_1951, kierstead_size_1953, okubo_diffusion_2001}. Here we just present one simple solution. Using the method of separation of variables we assume the existence of a solution of the form
\begin{equation}
 P(x, t) = X(x)\, T(t) \ .
 \label{eq:sep_anz}
 \end{equation}
Substituting this expression into (\ref{eq:KiSSeq}) we obtain the eigenvalue problem
\begin{eqnarray}
 T_n'(t) &=& \lambda_n T_n(t)  \nonumber\\[2mm]
D X_n''(x) + (\mu_0 - \lambda_n) X_n(x) &=& 0
\label{eq:KiSS_sepvar}
\end{eqnarray}
where the eigenfunctions $X_n(x)$ should satisfy the boundary conditions (\ref{eq:KiSSbc}), i.e. $X_n(0) = X_n(L) = 0$.
The solution to the last equation of system (\ref{eq:KiSS_sepvar}) is  well known
$$
X_n(x) = A_n \sin \sqrt{\frac{\mu_0 - \lambda_n}{D}} x + B_n \cos \sqrt{\frac{\mu_0 - \lambda_n}{D}} x  \ .
$$
It satisfies the boundary conditions only if $B_n = 0$ and the argument of the sine is divisible by $\pi n$ when $x = L$. From the last condition we obtain the eigenvalues
\begin{equation}
\lambda_n = \mu_0 - \frac{\pi^2 n^2 D}{L^2} \ , \qquad  n=1, 2,  \dots
\label{eq:KiSS_lmd}
\end{equation}

Solving the first equation of system (\ref{eq:KiSS_sepvar}), we find that $T_n(t) = C e^{\lambda_n t}$.
The problem is linear, therefore using the superposition principle, we can write $P(x,t)$ as the sum
$$
P(x, t) = \sum_{n=1}^\infty A_n \sin\left( \frac{n \pi x}{L} \right)
   e^{ \lambda_n t} \ ,
$$
where the coefficients $A_n$ depend on the initial conditions.

Thus, with the separation ansatz (\ref{eq:sep_anz}), we were able to present the dynamics of the full model as a sum of  simple ``modes''.
The dynamics of mode $n$ is determined by its eigenvalue $\lambda_n$. A mode grows if  $\lambda_n > 0$, decays if $\lambda_n < 0$, and can also oscillate if the imaginary part of $\lambda_n$ is not vanishing. Moreover, as $t\rightarrow \infty$ the full solution approaches to the mode that corresponds to the largest (principal) eigenvalue $\lambda^*$, because this mode has the fastest time exponent $T(t)\thicksim e^{ \lambda^*t}$. From (\ref{eq:KiSS_lmd}) we find that the largest eigenvalue $\lambda^* = \lambda_1$. Therefore,
$$
P(x, t)_{\overrightarrow{\ t\rightarrow \infty \ }}   A_1 \sin\left( \frac{\pi x}{L} \right)  e^{ \lambda_1 t} \ .
$$

The population grows, and so is able to persist, if $\lambda_1$ is positive.
By contrast, for $\lambda_1<0$ the population dies out exponentially on the whole habitat. Thus using expression (\ref{eq:KiSS_lmd}) for $n=1$, we obtain the fundamental relation between the diffusivity, the growth rate and the critical (minimal) size of the favourable patch, which provides the survival of the population
\begin{equation}
 L \geq L_0 = \pi \sqrt{ \frac{D}{\mu_0}}.
\label{eq:KiSS_Lc}
\end{equation}

Okubo \cite{okubo_horizontal_1978, okubo_diffusion_2001} demonstrated that this equation can be obtained by means of simple dimensional analysis. Suppose that $L_0 = f(D, \mu_0)$. Dimensionally, $[D] = m^2/s$, $[\mu_0] = 1/s$, and $[L] = m$. Thereby, the simplest combination of parameters yields
$$  L_0 = c \sqrt{\frac{D}{\mu_0}} \ , $$
where $c$ is a non-dimensional constant which equals $\pi$ in the 1D model. The same expression holds in 2D systems \cite{skellam_random_1951, kierstead_size_1953}, and it can be shown \cite{holmes_partial_1994, cantrell_spatial_2001} that the constant $c$ depends on the principal eigenvalue, thereby it represents the geometry of the model.

Equation (\ref{eq:KiSS_Lc}) can also be written in the form
\begin{equation}
D \leq D_0 = \frac{L^2 \mu_0}{\pi^2} \ , \label{eq:KiSS_Dc}
\end{equation}
yielding the critical, maximum mixing $D_0$ under which the population can survive.
Fig.~\ref{fig:skel}A and Fig.~\ref{fig:skel}B demonstrate the population
dynamics for different diffusivities. The population density
grows exponentially, if the diffusivity is smaller than $D_0$
(Fig.~\ref{fig:skel}A) and decays exponentially otherwise
(Fig.~\ref{fig:skel}B).

These results highlight the negative aspects of diffusivity for
a population and reveal an important ecological insight, namely that of a critical patch size. A finite population under the influence of random mixing must be larger than a minimal extension $L_0$ in order to sustain a stable population.
This critical size $L_0$ simply scales as the square root of the strength of mixing divided by the growth rate, (\ref{eq:KiSS_Lc}).
Despite the model's simplicity the fundamental results (\ref{eq:KiSS_Lc}) and (\ref{eq:KiSS_Dc}) have an important ecological message that prevails in more realistic settings.
If a favourable patch adjoins unfavourable areas, then the higher the diffusivity, the higher will be the loss rate across the boundaries, thus the larger should be the internal area and the growth rate on the habitat to compensate for these losses. Moreover, one large patch is better for the persistence of a species than two smaller patches of the same total size, since two patches would have four ends, that would double the loss rate. Diamond and May \cite{diamond_island_1976}, McMurtrie \cite{mcmurtrie_persistence_1978}, Cantrell and Cosner \cite{cantrell_effects_1998} applied this concept of a critical patch size to the design of national parks and natural reserves of optimal size and form.

\paragraph{Logistic growth}

As the growth of biomass in the KiSS model is not limited, one natural extension of the KiSS model is to consider a logistic growth function
\begin{equation}
\frac{\partial P(x, t) }{\partial t}
=  \mu_0 P \left( 1 - \frac{P}{K} \right)  +  D \frac{\partial^2 P}{\partial x^2}, \qquad \mathrm{for} \quad 0 \leq x \leq L
 \label{eq:KiSSloglim}
\end{equation}
and the same (hostile) boundary conditions (\ref{eq:KiSSbc}).
For this model only approximate solutions are available (see e.g. \cite{landahl_note_1959}, \cite{barakat_note_1959}, \cite{montroll_lectures_1968}, \cite{okubo_diffusion_2001}). The stationary solution can be expressed in terms of elliptic functions and was investigated by Skellam \cite{skellam_random_1951}, Levandowsky and White \cite{levandowsky_randomness_1977}, and Ludwig et al. \cite{ludwig_spatial_1979}. However, applying the invasibility criteria, we can conclude that this system possesses the same critical values as the KiSS model.
Indeed, as the population density approaches zero, equation (\ref{eq:KiSSloglim}) goes over to equation (\ref{eq:KiSSeq}). This means that if a population can invade in the KiSS model, it will also invade in the model with logistic growth, and vice versa.

\begin{figure}[tb]
\centerline{\includegraphics[height=5cm]{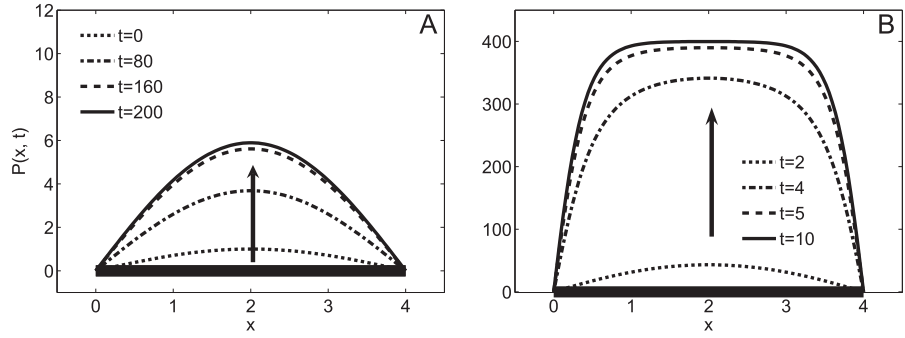}
}
\caption{Dynamics of the population with logistic growth, Eq.~(\ref{eq:KiSSloglim}), under different values of diffusion. (A) Conditions are nearly critical, $D = 3.2 \lesssim D_0$. (B) Conditions are sufficiently far from the survival-extinction transition $D = 0.1 \ll D_0$. Other parameters are $\mu_0=2$, $L=4$, $K = 400$, thus $D_0 = 3.24$.
The horizontal thick black line shows the location of the favourable patch.}
\label{fig:logist}
\end{figure}

Figs.~\ref{fig:logist} show examples of the population dynamics. In both figures, the resulting population density has a characteristic shape, with large densities in the central part of the patch and a decay of population numbers, the closer one comes to the hostile border.
Note, that if the conditions are close to the critical values (\ref{eq:KiSS_Lc}) or (\ref{eq:KiSS_Dc}), the growth of biomass becomes limited by the diffusive transport, and the maximum of density reached by the population can be much smaller than the carrying capacity. This is illustrated in Fig.~\ref{fig:logist}A where $\max (P(x))\approx 6$ even though $K = 400$. However, if the conditions are sufficiently far from the critical region, the carrying capacity is almost reached in the middle of the patch (Fig.~\ref{fig:logist}B).

To obtain an intuitive understanding into the influence of the density dependence in
equation (\ref{eq:KiSSloglim}) consider a KiSS model with an effective growth rate of $\mu^* = \left\langle \mu_0 (1 - P(x)/K) \right\rangle_x$, which equals the average growth rate of the logistically growing population.
The new effect now is that the effective growth rate $\mu^*$ decays with an increase of the population density $P(x)$. Therefore, also the maximal diffusivity $D_0^*$, admitting a survival of the population (\ref{eq:KiSS_Dc}), will be a decaying function of $P(x)$.
This means, however, that the biomass can grow only until the critical diffusivity is reached $D_0^*(P) = D$.
Thereby, if initially (when $P$ is negligible) $D$ is close to the
critical diffusivity of the KiSS model, a small increase of density is enough to decrease $D_0^*$ and to reach the balance between production and loss.

\paragraph{Finite mortality and other extensions}

A second unrealistic feature of the KiSS model is the assumption of infinite mortality outside of the favourable habitat. This assumption does not hold in many important cases and limits the applicability of the KiSS model.
For example, it hardly could be applied for phytoplankton simulations.
Ludwig et al. \cite{ludwig_spatial_1979} investigated an extension of the KiSS model, where this assumption of an absolutely hostile environment was relaxed.
In this study the authors assumed a positive growth rate $\mu_0$ inside and a finite mortality $m$ outside of the habitat
\begin{equation}
\mu(x) = \left\lbrace  \begin{array}{ll}
\quad \mu_0,  & \mbox{for } 0<x<L \\ \\
 -m,  &
\mbox{for } x \leq 0 \,\, \mathrm{or } \,\, x \geq L\ .
\end{array} \right.
\label{eq:LudwEvns}
\end{equation}

As the mortality is finite, the organisms can survive outside of the habitat and diffuse back.
This effect reduces the losses at the edges and the population can survive on a smaller favourable patch compared to the KiSS model. In the model (\ref{eq:LudwEvns})
the population can persist if
\begin{equation}
 L \ge 2 \sqrt{\frac{D}{\mu_0}}  \arctan \sqrt{ \frac{m}{\mu_0} }\ .
\label{eq:LudwEvnsCrit}
\end{equation}
As the mortality  approaches infinity  this value approaches the critical patch size $L_0$ of the KiSS model.

Many other examples of critical patch models can be found in the books by Okubo and Levin \cite{okubo_diffusion_2001} and Murray \cite{murray_mathematical_2003}. We just briefly mention some extensions. Okubo \cite{okubo_note_1972, okubo_advection-diffusion_1978} considered a model for growth and diffusion under an attractive force toward the centre of a patch. He found that the modified critical size equals
$ L_c = L_0 f\left( v^2/4 \alpha D \right)$,  where $f(0)=1$ and the function $f(  v^2/4 \alpha D )$ monotonically decreases with $v$ toward zero. Gurney and Nisbet \cite{gurney_regulation_1975} considered a model in which the growth rate parabolically depends on the distance from the habitat centre. This approach is more realistic since it includes a gradual transition from favourable to unfavourable areas.
Wroblewski et al. \cite{wroblewski_physical_1975}, Wroblewski and O'Brien \cite{wroblewski_spatial_1976} and Platt and Denman \cite{platt_general_1975} included the effect of grazing and obtained an expression similar to (\ref{eq:KiSS_Lc}) for the critical patch size, in which, however, $\mu_0$ is replaced by $\mu_0 - g$, where $g$ characterises the grazing rate.

\paragraph{Influence of boundary conditions and spatial arrangement}

\begin{figure}[tb]
 \centerline{\includegraphics[height=8cm]{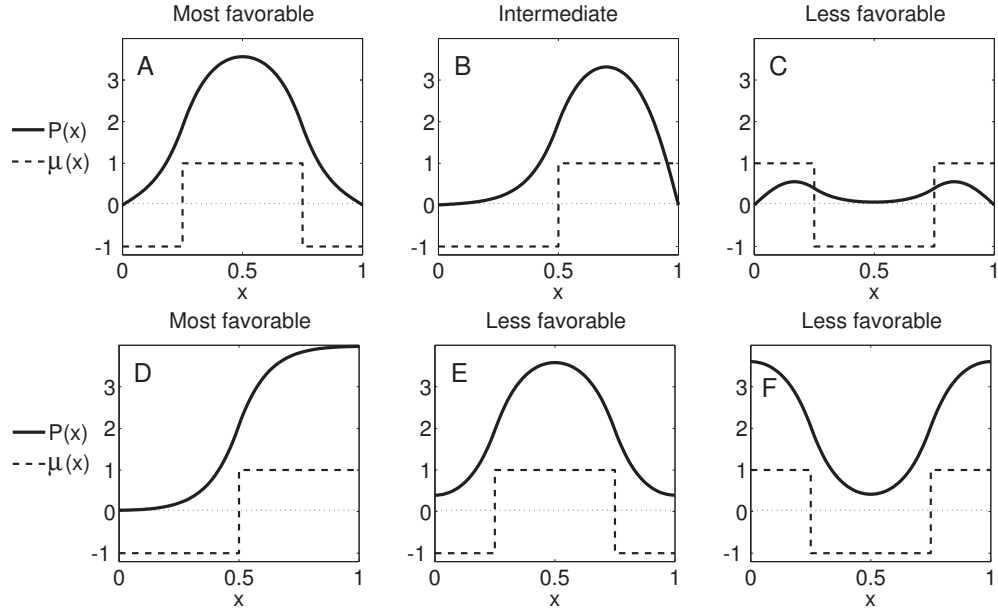}}
\caption{Comparison of several spatial arrangements of favourable ($\mu(x) = 1$) and unfavourable patches ($\mu(x) = -1$), assuming either lethal boundaries (top) or impenetrable boundaries (bottom)
at the borders of the habitat ($x=0$ and $x=1$) according to \cite{cantrell_effects_1991}.
Plotted is the final population density $P(x)$ in model (\ref{eq:rdiffNoAd}) assuming
logistic growth, $\mu(x,P)=\mu(x) (1-P/4)$  (solid line).
The patch quality, $\mu(x)$, is shown as dashed line.
Each case is classified either as ``most favourable'', ``intermediate'' or ``less favourable'', reflecting the total amount of biomass.}
\label{fig:arrang_leth}
\end{figure}

It is interesting to extend the analysis for more complicated spatial geometries. Seno \cite{seno_effect_1988} and Cantrell and Cosner \cite{cantrell_effects_1991} investigated the influence of a spatial sequence of favourable and unfavourable habitats and boundary conditions. Seno considered a population of organisms migrating between $n$
patches of different quality. Cantrell and Cosner \cite{cantrell_effects_1991} used a mean field representation of this problem.
They compared the total size of a population living on a finite habitat of size $L$, surrounded by either completely hostile environments or by impenetrable boundaries. Using the logistic model (\ref{eq:KiSSloglim}), they assumed that the habitat possesses patches of different quality, that is, the growth rate changes between positive or negative values, $\mu(x)=\pm 1$, provided that the favourable and unfavourable patches have the same total area (Fig.~\ref{fig:arrang_leth}).

In this model the best spatial configuration of favourable and unfavourable patches depends on the boundary conditions. If the exterior region is completely hostile then the location of a favourable patch in the middle of the habitat (Fig.~\ref{fig:arrang_leth}A) provides the best conditions for the population, as two buffer zones separate the favourable region from the completely hostile regions, which decreases the population losses. Therefore, the total biomass will be higher than that on the habitat shown in Fig.~\ref{fig:arrang_leth}B, which in turn is more preferable than the habitat shown in Fig.~\ref{fig:arrang_leth}C, where  two favourable patches adjoin the hostile surroundings.

If however, the boundaries act as a barrier, the maximum biomass is achieved if a favourable patch directly adjoins  one of the impenetrable boundaries (Fig.~\ref{fig:arrang_leth}D). Even though the derivation of this result requires rather sophisticated calculations, it is evident that in this configuration there is a single favourable patch which has only one border with an unfavourable environment, and so all possible losses are minimised. By contrast, any splitting of the favourable patch leads to a worsening of the habitat quality
(Fig.~\ref{fig:arrang_leth}E and \ref{fig:arrang_leth}F).

It is clear, that the choice of the best arrangement of habitat quality is very important for the design of national parks,  natural reserves, etc.
Further, the same mechanism also captures some aspects of relevance for the vertical distribution and competition of phytoplankton species.
In a water column the surface acts as an impenetrable barrier. Thus, in an
incompletely mixed water column, a more light limited or buoyant species obtains a competitive advantage to another species, whose favourable patch is located in subsurface layers \cite{huisman_changes_2004, ryabov_bistability}.

\paragraph{Arbitrary spatial dependence of the growth rate}

The behaviour of model (\ref{eq:rdiffNoAd}) in the case where the growth rate $\mu(x)$ is an arbitrary function of the coordinates was investigated by
Cantrell and Cosner \cite{cantrell_spatial_2001}.
Applying separation of variables to the linearised problem it is possible obtain a system of equations, similar to (\ref{eq:KiSS_sepvar})
\begin{eqnarray}
 T_n'(t) &=& \lambda_n T_n(t) \nonumber\\
\label{eq:KiSSGen_sepvar} \\
D X_n''(x) + \mu(x) X_n(x) &=&  \lambda_n X_n(x) \nonumber \ ,
\end{eqnarray}
where $X(x)$ should again satisfy certain boundary conditions.
Now, however, $\mu(x) \neq \mbox{const}$ and a solution to this equation is known only for some partial forms of the function $\mu(x)$. If the principal eigenvalue of these equations is positive, the stationary solution $P(x) = 0$ is not stable, implying that the population can invade and establish on the habitat. It can be shown that this is always the case if the average growth rate is positive, $\int \mu(x) dx > 0$. However, this condition is not necessary and does not hold even in the simple models considered before. General conditions of uniqueness and the existence of positive eigenvalues were obtained by Hess and Kato \cite{hess_periodic-parabolic_1991}, Senn and Hess \cite{senn_positive_1982}, Cantrell and Cosner \cite{cantrell_spatial_2001}.

The analysis of 2D and 3D models raises even more questions. In higher dimensions the persistence of a population may depend on the geometric form of the favourable patches \cite{diamond_island_1976}, the form of the edges separating the patches and finally it may depend on the behaviour of individuals, moving across or along the edges \cite{fagan_habitat_1999}.

\subsection{Persistence on an infinite habitat \label{sec:FK}}

\paragraph{Travelling fronts}
While in the previous section we have highlighted some negative aspects of diffusion for the fate of a population, in this section we show that depending on the circumstances diffusion may as well support population growth.
Maybe the most drastic example is the possibility to generate the spread or geographic expansion of a population into a new area.
Consider, for simplicity, an infinite homogeneous habitat which provides a positive growth rate $\mu_0$ everywhere. If, as it is commonly assumed (i.e., there is no Allee effect), the stationary state $P=0$ is unstable, the appearance of organisms in one spot will lead to their expansion over the whole habitat.
Assuming logistic growth the spatio-temporal dynamics of the population can be represented by the following equation
\begin{equation}
 \frac{\partial P(x, t)}{\partial t} = \mu_0 P \left(1 - \frac{P}{K}  \right) + D \frac{\partial^2 P}{\partial x^2} \ .
\label{eq:FK}
\end{equation}
%
which is known as the Fisher-Kolmogorov equation after Fisher
\cite{fisher_wave_1937}, who considered the logistic dynamics of advantageous
genes and Kolmogorov et al. \cite{kolmogorov_study_1937}, who investigated a
general form of this problem (see also \cite{murray_mathematical_2003,
kot_elements_2001, van_saarloos_front_2003} ).


\begin{figure}[tb]
\centerline{\includegraphics[height=4cm]{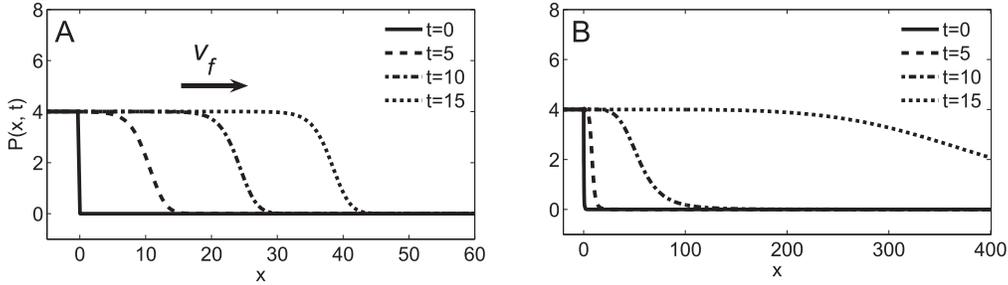}}
\caption{Front-propagation process  in the Fisher-Kolmogorov model (\ref{eq:FK}).
(A) Propagation of a travelling front, if initially $P(x, 0) = 0 \mbox{, for } x > 0$.
(B) Propagation of a ``pseudo-wave'' in the limit of zero diffusivity,  if initially $P(x, 0) = 1/(x + 1)^5$ for $x > 0$. In both plots $P(x, 0) = 1$ for $x \leq 0$.}
\label{fig:FK}
\end{figure}

Fisher \cite{fisher_wave_1937} and Kolmogorov et al. \cite{kolmogorov_study_1937} have shown that if initially some part of the habitat is not occupied, $P(x)=0$,
then the population will propagate into this part with the constant velocity
\begin{equation}
v_f = 2 \sqrt{\mu_0 D} \ .
\label{eq:FK_frontvel}
\end{equation}
This solution is illustrated in Fig.~\ref{fig:FK}A which shows a population propagating from left to right through a 1D habitat.
As will be shown below, the magnitude of the propagation velocity is a crucial factor not only for an invasion process but also for the survival of a population in the presence of drift, sinking or other advective processes \cite{dahmen_life_2000, speirs_population_2001, straube_mixing-induced_2007}.
There is a variety of ways to derive relation (\ref{eq:FK_frontvel}). The more common and rigorous approach suggests to assume a travelling solution of the form $P(x - v t)$ and then to prove that this solution is stable only if $v = v_f$  (see e.g. \cite{kot_elements_2001}). Below 
we will provide another heuristic derivation confirming the validity of this expression.

Note that in large aquatic basins the horizontal turbulent mixing increases with the scale of phenomena \cite{ozmidov_horizontal_1968, okubo_oceanic_1971, nihoul_modelling_1975, ozmidov_phytoplankton_1998}. Petrovskii \cite{petrovskii_diffusion_1999, petrovskii_plankton_1999} showed that this should result in the increase of the front propagation velocity. Moreover, this velocity should grow with the size of the area occupied by a population.

\paragraph{Pseudo waves}
We should stress a condition which is sometimes missed.
The velocity $v_f$ is the minimal possible propagation velocity, which is realised if the population invades into an empty area or if, at least, in this area $P(x, 0)$ decays more rapidly than a Gaussian distribution (see  Fig.~\ref{fig:FK}A).
Another pattern of so-called ``pseudo-waves'' may occur for special initial conditions if from the start a small amount of biomass is distributed over the whole space.
This is illustrated in Fig.~\ref{fig:FK}B where the initial distribution of biomass is algebraically decaying for $x>0$ and so is visually indistinguishable from that in Fig.~\ref{fig:FK}A. As shown this initial configuration yields a much faster wave-like spread of the population.
Strictly positive initial conditions lead to simultaneous logistic growth toward the carrying capacity at every position and, even in the limit of zero diffusivity, a wave-like pattern arises because the capacity is reached at slightly different times in different points and not due to the transport of biomass by diffusion.

Pseudo-waves appear if the time scale of diffusive transport, $\tau_D$, is slower than the time difference $\tau_{dem}$ of demographic processes in the neighbouring points
\begin{equation}
 \tau_D \sim \frac{\Delta x^2}{2 D} <
\tau_{dem} \sim - \frac{\Delta x  }{\mu(P) P } \mfrac{\partial P}{\partial x} \ ,
\label{eq:PseudoWCrit}
\end{equation}
where $\tau_{dem}$ can be found expanding the equation $P(x, t) = P(x + \Delta x, t + \tau_{dem})$ into a series. It is easy to check,
that for a Gaussian distribution of $P(x, 0)$ both sides of (\ref{eq:PseudoWCrit}) are proportional. In contrast, for algebraic or exponential distribution of $P(\Delta x, 0)$ the right hand side will be larger if $\Delta x$ exceeds a threshold value.
Consequently, the diffusive transport will be always slower than the time scale  $\tau_{dem}$ of the demographic process, giving rise to a ``pseudo-wave''.

\paragraph{Advection}

\begin{figure}[tb]
\centerline{\includegraphics[height=6cm]{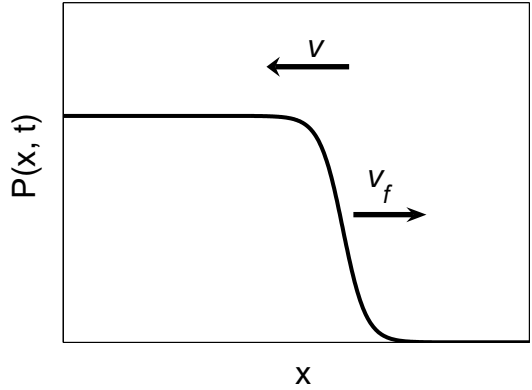}}
\caption{Schematic presentation illustrating the spread of a population into an oppositely directed flow of velocity $v$.
Initially the 1D habitat is populated on the left, while the right is not occupied by the species. In a system without advection the population would propagate into the right with the front velocity $v_f$, given by equation (\ref{eq:FK_frontvel}). This spread is hindered by the advective flow, which is aimed into the opposite direction. As a result
the population front propagates with the reduced velocity $v_f' = v_f - v$.
The population can persist only if $v_f \geq v$, otherwise it is washed out.  }
\label{fig:flowdir}
\end{figure}

Consider now an extension of the Fisher-Kolmogorov equation for a population which
is additionally subjected to an advective flow with constant velocity $v$
\cite{murray_minimum_1983, nelson_non-hermitian_1998, dahmen_life_2000, speirs_population_2001, straube_mixing-induced_2007, birch_bounding_2007, koszalka_plankton_2007}
\begin{eqnarray}
 \frac{\partial P(x, t)}{\partial t} = \mu_0 P \left( 1 - \frac{P}{K}\right)  -v \frac{\partial P}{\partial x} + D \frac{\partial^2 P}{\partial x^2}  \ .
\label{eq:FK_adv}
\end{eqnarray}
To investigate the role of advection let us again suppose initial conditions, such that only the left part of the habitat is populated, while the remaining part is not occupied by the species (see Fig.~\ref{fig:flowdir}). In the absence of advection in this system, the population would propagate into the right with the velocity $v_f$.
Now assume that this spread occurs in an advective flow, which is going into the opposite direction with the velocity $v$.
This can be analysed best by considering a frame of reference moving parallel to the flow with the same velocity $v$, so that in this frame the flow velocity is zero. In the moving reference frame the population dynamics obey equation (\ref{eq:FK}) and the front propagates with the velocity $v_f$ given by (\ref{eq:FK_frontvel}). Therefore, in the fixed reference frame, the propagation velocity is reduced by the velocity $v$ of the advective flow, so that the front propagates with the velocity $v_f' = v_f - v$.

Obviously, the sign of the reduced propagation velocity determines the persistence of the species. A negative propagation velocity leads to a population wash-out, whereas a positive propagation velocity results in the invasion of the empty habitat. Thus, on an infinite habitat in an advective flow the population can survive only if
\begin{equation}
v \leq  v_f = 2 \sqrt{\mu_0 D}\ ,
\label{eq:FK_v_max_front}
\end{equation}
or, written in a different form
\begin{equation}
D \geq D_{min} = \frac{v^2}{4 \mu_0} \ ,
\label{eq:FK_crit_diff_adv}
\end{equation}
which means that in a flow a population can persist only if the diffusivity exceeds the threshold value $D_{min}$.
Equations (\ref{eq:FK_v_max_front}) and (\ref{eq:FK_crit_diff_adv})
relate the critical values of diffusivity $D$ and flow velocity $v$
and illustrate the constructive interplay that can arise in the presence of both processes, advection and diffusion.
The region in the $(v, D)$-coordinate plane, where a population can outgrow an advective flow, is visualized in Fig.~\ref{fig:semi_circle} (area under the solid line).
Note that if the habitat is unlimited then for any flow velocity $v$  there is a diffusivity which  provides the survival of the population.
This transition plays an important role in many ecological situations and constitutes, for example, a necessary condition for the persistence of a population in a river \cite{speirs_population_2001}, as well as for the persistence of sinking phytoplankton species in a vertical water column \cite{riley_quantitative_1949, shigesada_analysis_1981, huisman_do_2002}.

\paragraph{Derivation of the front propagation velocity}

These results about the spread of a population in an advective flow can be elegantly used to derive the population spread, Eq.~(\ref{eq:FK_frontvel}), in a system without advection.
For these aims we consider the general model
\begin{equation}
\frac{\partial P(x, t)}{\partial t} = \mu(x) P -v \frac{\partial P}{\partial x} + D \frac{\partial^2 P}{\partial x^2}  \ .
\label{eq:FK_linear}
\end{equation}
Here, the local growth term has been linearised for small densities, however we allow
for an arbitrary spatial dependence of the growth rate, $\mu=\mu(x)$.
This model can be simplified with the following transformation
%
\begin{equation}
P(x, t) = \exp \left(  \frac{v x}{2D} \right) \, \tilde{P} (x, t) \  .
\label{eq:AdvecTrans1}
\end{equation}
The aim of this `trick' is to eliminate the advective term and as a result we
obtain the following equation of a system without advection, however with a modified growth term
\begin{equation}
\frac{\partial \tilde{P}(x, t)}{\partial t} =\left(  \mu(x) -  \frac{v^2}{4 D} \right)  \tilde{P}  + D \frac{\partial^2 \tilde{P}}{\partial x^2}  \ .
\label{eq:FK_transformed}
\end{equation}
Note that the transformation (\ref{eq:AdvecTrans1}) should be applied also to any boundary conditions of the original system and may alter them. However, if we can neglect the influence of boundaries (e.g. if the habitat is infinite, but still heterogeneous) advection obtains a very simple ecological interpretation as an additional mortality of strength $v^2/4D$.
This means that the presence of advection effectively reduces the growth rate $\mu(x)$,
and this effect increases as $D \rightarrow 0$.

Now we can ``derive'' the velocity $v_f$ of the front propagation in the Fisher-Kolmogorov equation. On a homogeneous infinite habitat ($\mu(x) = \mu_0$) the population can survive only if the growth rate is positive. Thus, using (\ref{eq:FK_transformed}) the velocity should be constrained
$$
v \leq 2 \sqrt{\mu_0 D} .
$$
However, as we showed before, the maximal possible advection velocity, $v$, is equal to the front propagation velocity, $v_f$, which finally leads us to formula  (\ref{eq:FK_frontvel}).

\subsection{Finite habitats in an advective flow: the ``drift paradox'' \label{Seq:DriftPar}}

While the Fisher-Kolmogorov equation (\ref{eq:FK}) assumes an infinite homogeneous habitat,
such conditions are a strong idealisation for most natural populations.
To describe some more realistic situations, in the following we investigate the role of advection for a population in a heterogeneous environment which is additionally constrained by certain boundary conditions.

\paragraph{Separation of variables \label{sec:sep_var}}

Consider again the general model
(\ref{eq:FK_linear}) of the previous section, which now is supposed to be complemented by some boundary conditions. Again we use the change of variables (\ref{eq:AdvecTrans1}) to eliminate the advection term.
The linear form of equation (\ref{eq:FK_transformed}) allows a separation of variables
$ \tilde{P}(x, t) = X(x)\, T(t) $
and by comparison to Eq.~(\ref{eq:KiSSGen_sepvar}) we obtain the eigenvalue problem for the time-independent eigenfunctions
\begin{equation}
D X''(x)  + \mu(x) X(x) = \left( \lambda^v + \frac{v^2}{4D} \right)  X(x) \ ,
\label{eq:ev_prob}
\end{equation}
while the equation for $T(t)$ remains unchanged.
Here, $\lambda^{v}$ denotes the eigenvalues for the problem in the presence of advection.
Furthermore, $X(x)$ should satisfy certain boundary conditions, which depend on the boundary conditions obtained for $\tilde{P}(x, t)$.
If we introduce $\lambda =  \lambda^v + v^2/4D$ this equation will take the form of the second equation of system (\ref{eq:KiSSGen_sepvar}) for the problem without flux. Thus we can conclude that the presence of advection for a population under boundary conditions
simply reduces the eigenvalues \cite{grassberger_diffusion_1982, nelson_non-hermitian_1998, dahmen_life_2000}
$$
\lambda^v = \lambda - \frac{v^2}{4D} \ .
$$

With the same arguments as in Section~\ref{sec:crit}, to provide the persistence of a population, the largest eigenvalue $\lambda^v$ must be positive. Therefore, a population is only able to persist in a flow if the proper model without flow has the principal eigenvalue
\begin{equation}
\lambda^* \geq \frac{v^2}{4 D} \ .
\label{eq:KiSS_lmd_adv}
\end{equation}

\paragraph{KiSS model with advection}

\begin{figure}[tb]
\centerline{\includegraphics[height=7cm]{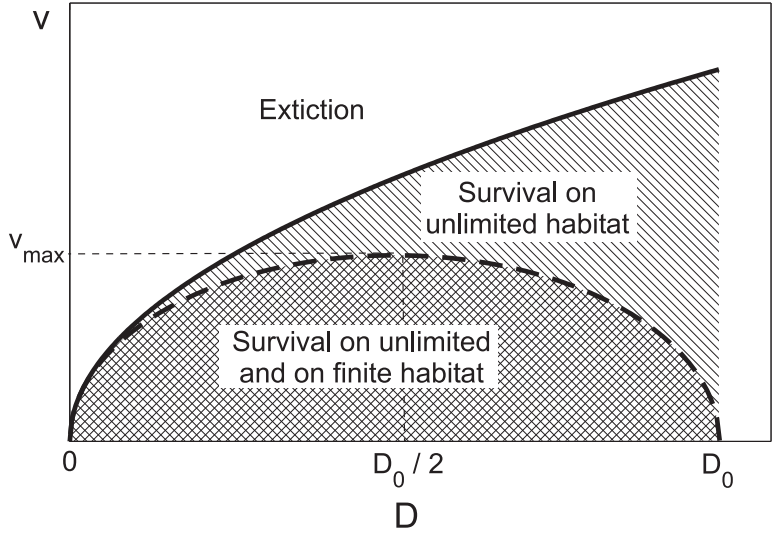}}
\caption{Parameter regions which permit the persistence of a population in an advective flow.
Persistence of the population is possible below the critical curves in the $(D,v)$-parameter plane. Two different scenarios are compared. The solid line shows the persistence regime
for an infinite uniform habitat (cross and diagonal hatching). In this case
the limiting velocity scales as the square root of diffusivity $v~\sim \sqrt{D}$, equation (\ref{eq:FK_v_max_front}).
In comparison, the dashed line shows the result for the persistence of a population on a finite favourable patch with lethal boundaries (cross hatching). Here, the persistence regime yields a semi-circle in the parameter plane, equation (\ref{eq:KiSS_D_Bound}).
Assuming intermediate growth conditions, e.g. similar to the model by Ludwig et al. (\ref{eq:LudwEvns}), one should expect the critical curve to be located somewhere  in-between the solid and the dashed line.}
\label{fig:semi_circle}
\end{figure}

As a simplest example, consider the KiSS model (\ref{eq:KiSSeq}) in the presence of an  advective flow. Recall, that this model describes a population on a finite favourable patch provided that the population density vanishes at the boundaries. These boundary conditions ($P(0, t) = P(L, t) = 0$) are not changed by the coordinate transform (\ref{eq:AdvecTrans1}), so that the functions $\tilde{P}(x, t)$ and $X(x)$ also must vanish at the boundaries. Therefore, we will obtain the same expression (\ref{eq:KiSS_lmd}) for the eigenvalue spectrum with the dominant eigenvalue $\lambda^* = \mu_0 - \pi^2 D/L^2$.
Taking into account (\ref{eq:KiSS_lmd_adv}), we can easily derive the conditions for the persistence
of a population on a finite favourable patch in an advective flow
\begin{equation}
v \leq  v_{f} =2 \sqrt{D \lambda^*} = 2 \sqrt{D \left( \mu_0 - \frac{\pi^2 D}{L^2}\right) }\ .
\label{eq:KiSS_vc}
\end{equation}
This expression describes a semi-circle in the $(D,v)$ parameter plane
(see dashed line in Fig.~\ref{fig:semi_circle}). Note that as $L$ approaches infinity or $D$ approaches zero, the maximal velocity (\ref{eq:KiSS_vc}) goes over to the condition (\ref{eq:FK_v_max_front}). This means that in the limit of small diffusivity ($D\ll \mu L^2/\pi^2$) the conditions for survival on a finite patch are almost the same as those on a homogeneous infinite habitat. Thus, for small $D$ the two critical curves in Fig.~\ref{fig:semi_circle} almost coincide and show a similar behaviour.
For larger diffusivity, however, these curves diverge because a finite habitat provides a slower propagation velocity (\ref{eq:KiSS_vc}) than an infinite habitat  (\ref{eq:FK_v_max_front}).
Finally, for large values of $D$ the increase of the losses across the habitat edges results in the upper diffusivity limit $D_{max}$ on a finite habitat.

Solving this inequality for $D$, we obtain a limiting interval $D_{min}(v) \leq D \leq D_{max}(v)$, with
\begin{equation}
D_{min/max} = \frac{D_0}{2}  \left(1 \pm  \sqrt{1 - \frac{\pi^2 v^2}{L^2 \mu_0^2}} \right) \ .
\label{eq:KiSS_D_Bound}
\end{equation}
Here, $D_0$ is the critical (maximal) diffusivity (\ref{eq:KiSS_Dc}) in the KiSS model. The interval $\left[ D_{min}, D_{max}\right] $ specifies the limits of  mixing intensity which prevent the population wash-out ($D>D_{min}$), but still enable the persistence of the population on a finite patch ($D<D_{max}$). These values are real only if
\begin{equation}
v \leq v_{max} = \frac{L \mu}{\pi} \ .
\label{eq:vmax_cond}
\end{equation}
Note that the critical velocity $v_{max}$ is a threshold when the characteristic time scale of growth $\tau_{gr} = 1/\mu$ becomes slower than the time scale of advection $\tau_{v} = L/v$. If $v>v_{max}$ the population cannot persist, because on the one hand the large advection requires a strong mixing intensity to provide the expansion of organisms upstream, but on the other hand such mixing increases the transport of organisms into unfavourable areas and the population becomes extinct.

\paragraph{Critical patch size}
In the following we investigate the influence of advection on the critical patch size in the KiSS model (\ref{eq:KiSSeq}) and in the model by Ludwig et al. (\ref{eq:LudwEvns}). In the first model the population vanishes at the habitat edges, whereas in the second the population density should vanish when $x \rightarrow \pm \infty$.  Therefore, in both models the transformation (\ref{eq:AdvecTrans1}) does not alter the boundary conditions and the presence of an advective flow simply reduces the growth rate $\mu$. This leads to the effective growth rate
$$
\mu_0^v = \mu_0 - \frac{v^2}{4D}
$$
in the KiSS model and to
$$
\mu^v(x) = \left\lbrace  \begin{array}{ll}
\quad \mu_0 -  \mfrac{v^2}{4D}\ ,  & \mbox{for } 0<x<L \\ \\
 - \left(  m + \mfrac{v^2}{4D} \right) \ ,  &
\mbox{for } x \leq 0 \,\, \mathrm{or } \,\, x \geq L\ ,
\end{array} \right.
$$
in the model by Ludwig et al.

Substituting the modified growth rate $\mu^v$ into equation (\ref{eq:FK_transformed}), we find similar expressions for the critical patch size as in Eqs.~(\ref{eq:KiSS_Lc}) and  (\ref{eq:LudwEvnsCrit}), where $\mu$ should be replaced by $\mu^v$.
Thus,  for the KiSS model in an advective flow the critical patch size equals
\begin{equation}
L_0^v =  \pi \sqrt{\mfrac{D}{\mu_0^v }} =  \pi \sqrt{\dfrac{D}{\mu_0 -  v^2/4D}}  =    \dfrac{2 \pi D}{\sqrt{v_f^2 -  v^2}}\ ,
\label{eq:KiSS_Lc_adv}
\end{equation}
and for the model by Ludwig et al. we obtain
\begin{equation}
 L^v =  \frac{4 D}{\sqrt{v_f^2 -  v^2}} \arctan \sqrt{\frac{4 m D + v^2}{v_f^2 -  v^2}} \ ,
\label{eq:Ludw_Lc_adv}
\end{equation}
where $v_f = 2 \sqrt{D \mu_0}$. Note that both $L_0^v$ and $L^v$ approach infinity as $v \rightarrow v_f$. In particular, this can happen if $D \rightarrow D_{min} = v^2/4\mu_0$ which is of importance in marine biology as climate models predict that the ongoing global warming may result in a higher stratification of the ocean water \cite{bopp_potential_2001, sarmiento_response_2004}, increasing thereby the requirement on the critical (vertical) patch size for sinking phytoplankton species.

\paragraph{Persistence in a river}
Speirs and Gurney \cite{speirs_population_2001} investigated the conditions for the population survival in a river of length $L$ and with a flow velocity $v$.
This problem is known as the ``drift paradox'' because any advection will ensure that the average location of a population will move downstream, so at first glance it seems counter-intuitive that a population can persist in a river \cite{hershey_stable_1993}.
In their study Speirs and Gurney used the linear model (\ref{eq:FK_linear}) and assumed an impenetrable  boundary upstream and a totally hostile environment downstream. In this model the critical size $L^v$ of the favourable patch depends  on the ratio between the flux velocity $v$ and the front propagation velocity $v_f$
$$
L^v = \frac{2 D}{\sqrt{v_f^2 -  v^2}}  \arctan   \sqrt{\frac{v_f^2 - v^2}{v^2}} \ .
$$
Note that $L^v$ again increases with an increase of the advection velocity $v$ and approaches infinity as $v \rightarrow v_f$. Furthermore, because the favourable patch in this model has only one boundary with the hostile environment, we obtain the limit $L^v \rightarrow L_0/2$ as $v \rightarrow 0$,
where $L_0$ is the critical patch size  (\ref{eq:KiSS_Lc}) of the KiSS model.

Extending this model, Pachepsky et al. \cite{pachepsky_persistence_2005} derived conditions for the persistence and spread of a population of organisms living and reproducing on the sediment and occasionally entering the water flow where they can drift and disperse.

\paragraph{Locally elevated growth rate}
Dahmen et al. \cite{dahmen_life_2000} considered another extension of the model by Ludwig et al. (\ref{eq:LudwEvns}), assuming an advective flow and periodic boundary conditions. Furthermore they suggested a simple experimental set-up in which a light limited colony of bacteria grows on a ring. Almost the whole ring is shaded and unfavourable for the population. Only a small area is illuminated through a window of length $L$, which moves around the ring with a constant velocity $v$. Up to a change of  reference frame, this set-up is equivalent to an advective flow and, changing the speed $v$, one can easily regulate the ``advection'' rate.

\begin{figure}[tb]
 \centerline{\includegraphics[height=6cm]{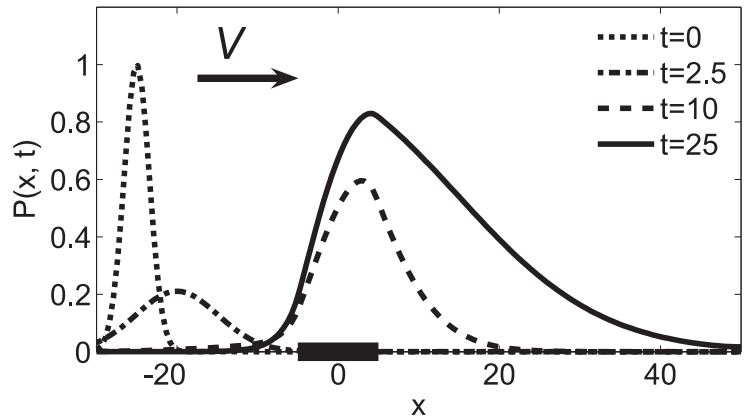}}
\caption{Locally elevated growth rate provides the persistence of a population in an advective flow with velocity $v$, Dahmen et al.  \cite{dahmen_life_2000}. A favourable patch ($\mu(x) = \mu_0 > 0$) is indicated by the horizontal thick line.
}
 \label{fig:Dahm}
\end{figure}

Depending on the parameters  and the velocity of the light supply, the population can become extinct, can be localised (the maximum of density tracks the location of the favourable patch), or delocalised \cite{nelson_non-hermitian_1998} (due to the periodic boundary conditions the population can survive even if the advection velocity $v$ exceeds the propagation velocity $v_f$ provided that the average growth rate is positive). Fig.~\ref{fig:Dahm} shows an example for such a localisation.

Solving the eigenvalue problem Dahmen et al. exploited a similarity of equation (\ref{eq:ev_prob}) to the well studied ``square well potential'' problem of quantum mechanics  \cite{landau_quantum_1981}.
They showed that depending on the dimensionless parameter $ \overline{x} = (2/L) \sqrt{D/(\mu_0 + m)}$, which characterises the growth rate in relation to the diffusivity and the habitat length, the critical flow velocity can be expressed as
\begin{equation}
\begin{array}{llll}
 v_c &=& 2 \sqrt{D \left( \mu_0 - \mfrac{D \pi^2}{L^2}  \right) }\ , \qquad &\mbox{ if } \overline{x} \ll 1 \\
\\
 v_c &=& 2D \sqrt{\left( \left[ \mfrac{(\mu_0 + m) L }{2 D} \right]^2 - \mfrac{m}{D}  \right) }\ , \qquad &\mbox{ if } \overline{x} \gg 1 \\
\end{array}
\label{eq:Dahmen}
\end{equation}
where all parameters have the same meaning as in (\ref{eq:LudwEvns}).
The first solution also corresponds to the limit of high mortality ($m \rightarrow \infty$) and coincides with equation (\ref{eq:KiSS_vc}), which was derived for the KiSS model with advection. The second solution describes a strongly mixed system. If the advection velocity is higher, the population becomes extinct. The value of the critical patch size and the critical diffusivity can be easily expressed from equation (\ref{eq:Dahmen}).

Examining bacterial growth,  Lin et al. \cite{lin_localization_2004} confirmed the results of Dahmen et al. \cite{dahmen_life_2000} experimentally and by means of numerical simulation.
Joo and Lebowitz \cite{joo_population_2005} carried out computer simulations in a stochastic spatially discrete population model and obtained similar results, confirming the robustness of the model.

\paragraph{Locally elevated diffusivity}

\begin{figure}[tb]
 \centerline{\includegraphics[height=6cm]{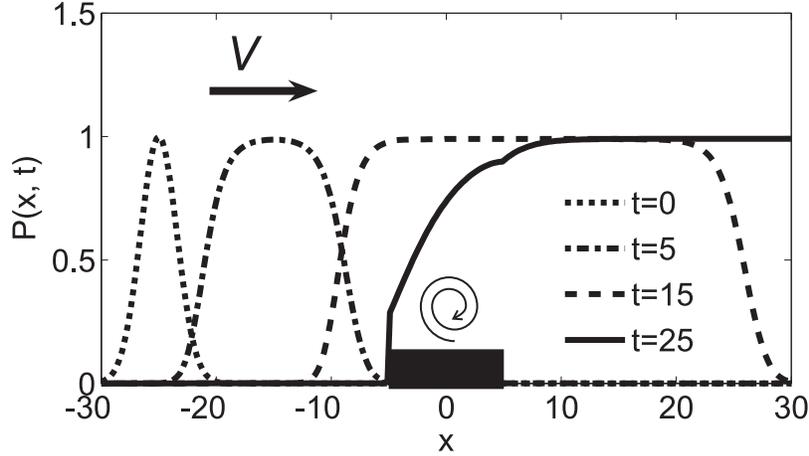}}
 \caption{Locally increased diffusivity stops the population wash-out in an advective flow with velocity $v$, according to Straube and Pikovsky \cite{straube_mixing-induced_2007}.
The horizontal thick black line indicates the location of the patch with increased diffusivity $D_1$.
}
 \label{fig:Pik}
\end{figure}

Consider a population growing on an infinite favourable patch in an advective flow. If the diffusion is too low this patch will not provide a proper propagation velocity (\ref{eq:FK_frontvel}) and the population will become extinct. Straube and Pikovsky \cite{straube_mixing-induced_2007} noted that locally increased diffusivity can drastically change the situation. A sufficiently large and well mixed patch can stop the drift of biomass, stabilising the population dynamics downstream (Fig.~\ref{fig:Pik}). Straube and Pikovsky considered the Fisher-Kolmogorov equation with advection (\ref{eq:FK_adv}), assuming that a patch of length $L$ has an elevated level of diffusivity $D_1$, which exceeds the diffusivity $D$ in the rest of the habitat. Note that if the diffusivity depends on the coordinate, $D=D(x)$, equation (\ref{eq:FK_adv}) takes the form
$$
 \frac{\partial P(x, t)}{\partial t} = \mu_0 P  \left(  1 - \frac{P}{K}\right)  -v \frac{\partial P}{\partial x} +
 \frac{\partial }{\partial x} \left[ D(x) \frac{\partial P}{\partial x} \right]  \ .
$$
Straube and Pikovsky \cite{straube_mixing-induced_2007} showed that the population survives,
if the size of the intensively mixed patch is larger than
$$
L_{cr} = \frac{2 D_{1}}{\sqrt{4 \mu_0 D_{1} - v^2}} \arctan \sqrt{\frac{v^2 - v^2_f}{4 \mu_0 D_{1} - v^2}} \ .
$$
As the diffusivity $D_1$ approaches infinity the critical patch size approaches
$$
\lim_{D_1 \rightarrow \infty} L_{cr} =  \frac{\sqrt{v^2 - v_f^2}}{2 \mu_0} \ .
$$
Thus, even for the infinite mixing intensity the critical patch should be of finite size.

\paragraph{Techniques from quantum mechanics}

Birch et al. \cite{birch_bounding_2007} considered the Fisher-Kolmogorov equation with variable growth rate and advection on a 2D plane.
In this study they made use of the structural similarity between equation (\ref{eq:ev_prob}) and the time-independent Schr\"{o}dinger equation.
Based on this similarity, they demonstrated a few examples where the
application of perturbation theory, the method of Wentzel, Kramers, and Brillouin (WKB), and other quantum mechanical techniques are beneficial for the analysis of equations similar to (\ref{eq:ev_prob}).
In particular, Birch et al. were able to determine a trade-off between the critical diffusivity and growth rate, which provide the persistence of the population in this system.
We believe that the application of such well-known techniques from quantum mechanics has not been fully exhausted yet. Such methods could be a promising direction for the further development of the theory of extended populations.

\paragraph{Summary} At the end of this section we want to summarize the main results.
On an infinite homogeneous patch a stationary state is achieved when the growth rate is equal to the mortality. On a finite patch, which adjoins some unfavourable environment, the growth rate should exceed mortality to compensate for losses across the patch edges. The internal patch production grows with the patch size and with the growth rate, whereas the losses at the edges increase with the diffusivity. The spread of organisms over a habitat is defined by the propagation velocity of the front of the population density. This velocity scales as a square root of the growth rate and diffusivity. A population can survive in a flow only if the propagation velocity is higher than the velocity of the flow.

Thus, an increase of the growth rate improves the conditions for the survival on a finite favourable patch and in a flow. The increase of the patch size increases the internal patch production, whereas the losses across the patch edges show only a weak dependence. The critical patch size is achieved, when the internal production is balanced by the external losses. The size of a large habitat has a small influence of the propagation velocity. However, an unfavourable environment around a small favourable patch truncates the propagating front, this can essentially reduce the propagation velocity and worsen the conditions in a flow. Finally, an increase of diffusivity on a finite habitat increases the losses into the unfavourable environment, which can lead to the population extinction. However, a minimal level of diffusivity is necessary to prevent the population wash-out in a flow. These two opposing processes result in a diffusivity window, which provides the population survival on a finite habitat. This window exists only if the flow velocity is less than some critical value.

\setcounter{equation}{0}
\section{Vertical phytoplankton distribution}
In the previous chapter, using simple one-dimensional models, we discussed the influence of mixing and advection on the population survival in a heterogeneous environment.
To model a non-uniform environment, we simply used an explicit form for the growth rate,
$\mu(x)$, however we did not discuss the origin of this heterogeneity. In this chapter we
extend these results by considering more complex models which describe resource limited population growth.
These models are based on simple physical and biological laws and consistently describe the dynamics of a population and its limiting resources. Thus, they demonstrate natural mechanisms which lead to the appearance of favourable patches. The intention of this chapter is twofold: on the one hand we aim to illustrate how the main findings of the previous chapter apply to the context of consumer-resource models.
On the other hand we show new effects arising due to new properties, which are not present in the simple models.

The main object, which we will use for illustration, is the dynamics of a vertical phytoplankton distribution.
This is important as phytoplankton are the primary producers in almost all aquatic food webs with a major influence on nearly all freshwater and marine ecosystems.
The two main factors limiting the production of phytoplankton are the availability of nutrients and light.  To understand how these  resources affect the phytoplankton biomass, consider their distributions in a water column. In general, the light intensity reduces with depth and, in nutrient rich regions of the ocean, the well illuminated  surface layer constitutes a favourable area for photosynthetic phytoplankton species. By contrast, the nutrient concentration can behave just in the opposing way. The sedimentation of dead biomass (detritus), with the successive remineralisation in the deep layers or in the sediment \cite{ebenhoh_benthic_1995} causes an increase of the nutrient concentration with depth \cite{yakushev_analysis_2007}. Thus, while light limitation may lead to the formation of a surface phytoplankton maximum, a lower nutrient concentration favours a phytoplankton build-up in deeper layers.  This tension between light and nutrient limitation from two opposite sides frequently causes optimal growth conditions in subsurface layers (e.g. \cite{abbott_mixing_1984, cullen_deep_1982, holm-hansen_deep_2004}). This fact often leads to the appearance of maxima of chlorophyll or biomass distributions at approximately 30-100 m depth. So called deep chlorophyll maxima (DCM) \cite{abbott_mixing_1984, venrick_deep_1973, cullen_deep_1982, venrick_phytoplankton_1993, holm-hansen_deep_2004} and deep biomass maxima (DBM) \cite{kononen_development_2003, bishop_transmissometer_1999} are ubiquitous phenomena and can be observed in many oligotrophic regions in the ocean, marine systems, and deep lakes.


Another important component of a stratified water column is an upper mixed layer (UML). A UML commonly occurs in oceans and lakes due to mechanical perturbation of the surface waters (e.g. due to wind, waves, and storms). This layer is separated from the deep layers by a thermocline  \cite{condie_influence_1997}, which is defined as a relatively thin layer below a UML characterised by an strong change in temperature with depth. Mixing in a UML is much stronger than in the layers below it. As a result, the distributions of nutrients, temperature, salinity, etc. are nearly uniform in a UML and have gradients below it. The depths of a UML can usually vary from 10~m to 100~m, see e.g. \cite{venrick_phytoplankton_1993}.


Compared to the models of the previous chapter, the system behaviour in a water column
can be further complicated due to a feedback loop between the biomass and resource distributions. The growing biomass shades light, consumes nutrients and is remineralised, which ultimately changes the total resource distribution. This, in turn, can lead to a new biomass distribution, which will generate a new resource profile and so on. As will be shown below, these complicated, self-organised dynamics can lead to new phenomena and diverse behaviour. For example, if the mixing is small, the final solution becomes non-stationary and oscillates \cite{huisman_reduced_2006}, whereas  in the presence of an upper mixed layer the system may exhibit bistability and the solution may be sensitive to the initial conditions \cite{yoshiyama_catastrophic_2002, ryabov_bistability}.

\paragraph{Equation of growth}
To formulate a mathematical framework for this chapter, let us consider a vertical water column of depth $Z_B$. Let $P(z, t)$ denote the density of phytoplankton at time $t$ and depth $z$. Note that $z = 0$ denotes the sea level surface and the $z$-axis is directed downward. For the sake of simplicity, assume that phytoplankton growth is limited only by the availability of light and a nutrient (the model can easily be extended to take into account multiple nutrient limitation \cite{grover_resource_1997}).
In our approximation the dynamics of a phytoplankton population obey a reaction-diffusion-advection equation, similar to equation (\ref{eq:rdiff}) considered in the previous chapter (see \cite{radach_vertical_1975, shigesada_analysis_1981, klausmeier_algal_2001, huisman_reduced_2006} among others)
\begin{equation}
\frac{\partial P(z, t) }{\partial t} = \mu(N, I)  P - m P - v \frac{\partial P }{\partial z} + \frac{\partial  }{\partial z} D \frac{\partial P}{\partial z} \ ,
 \label{eq:phyto}
\end{equation}
where $m$ is the mortality (compare to equation (\ref{eq:NP})), $v$ is the phytoplankton sinking velocity, and $D$ is the diffusivity, which in general can depend on $z$.

Furthermore, the growth rate $\mu(N, I)$ depends on the local values of light intensity $I(z, t)$ and nutrient concentration $N(z, t)$ at each vertical position.
If both nutrients are essential, $\mu(N, I)$
can be represented in the form of Liebig's law of minima
%
\begin{equation}
\mu(N, I) = \mu_0 \min \left[ f_I(I), f_N(N) \right] \ .
\label{eq:phyto_growth_min}
\end{equation}
It can be also written in the multiplicative form
\begin{equation}
\mu(N, I) = \mu_0 \, f_I(I) \, f_N(N) \ .
\label{eq:phyto_growth_mul}
\end{equation}
Here $\mu_0$ is the maximum growth rate and $f_I(I)$ and $f_N(N)$ describe the limitation by light and the nutrient. The specific form of these functions depends on many factors, for instance, strong light may photoinhibit photosynthesis and reduce the growth rate for large values of $I$ \cite{platt_photoinhibition_1980, geider_dynamic_1997, reynolds_modelling_1997}. However, usually it is suggested that $f(x) \rightarrow 1$  as $x \rightarrow \infty$, that is, the maximum growth rate is achieved when all resources are unlimited. For phytoplankton modelling, the most frequently used form is the Monod (or Michaelis-Menten) kinetics \cite{turpin_d.h._physiological_1992}
\begin{equation}
f_I(I) = \frac{I}{H_I + I} \ , \qquad f_N(N) = \frac{N}{H_N + N} \ ,
\label{eq:Menten}
\end{equation}
where $H_I$ and $H_N$ are the half-saturation constants for nutrient-limited and light-limited growth, respectively. However, this non-linear form often admits only numerical investigation. Analytical solutions are commonly possible only for a linear or algebraic form of $f(x)$ \cite{shigesada_analysis_1981, ebert_critical_2001}.

\paragraph{Boundary conditions} By default, we assume that the surface and bottom are impenetrable for phytoplankton
\begin{equation}
  \left. \left(  v P(z, t) - D \frac{\partial P}{\partial z} \right) \right|_{z = 0, Z_B} = 0\ .
\label{eq:boundary_phyto}
\end{equation}

To model a stratified water column, one can either separately solve the equations in a UML and below it, supposing infinite mixing within the UML and a small diffusivity $D_D$ in deep layers. Assuming continuity of the flux across the thermocline, we obtain the boundary condition at  the bottom of a UML (see e.g. \cite{huisman_population_2002})
$$
 \left. v P(z) \right|_{z = Z_{T} - 0}= \left. \left(  v P(z) - D_D \frac{\partial P}{\partial z} \right) \right|_{z = Z_{T} + 0} \ ,
$$
where $Z_{T}$ is the depth of the thermocline. On the other hand, to simulate the water column in a single framework, one can assume a gradual transition from  a UML to the deep layers \cite{ryabov_bistability}
\begin{equation}
D(z) = D_D + \mfrac{D_U  - D_D}{1 + e^{(z-Z_{T})/w}} \ ,
\label{eq:Dz}
\end{equation}
where $D_U$ and $D_D$ are the diffusivities within and below a UML,  respectively, and the parameter $w$ characterizes the width of the thermocline.

So far, we did not specify any equations for the distribution of nutrients $N(z,t)$ and light $I(z,t)$.
In the next two sections we will review several models which couple the light and nutrient dynamics with phytoplankton growth (\ref{eq:phyto}). First we will consider models in which phytoplankton growth is limited only by the light availability, whereas in a second step, we will review more complex models, incorporating both light and nutrient limitation.

\subsection{Light limitation}

In this section we will consider theoretical models in which the light gradient is a key factor.
Such models give an adequate description for eutrophic aquatic environments (as observed in many regions), where the nutrients are in ample supply and  light becomes a crucial factor which determines the distribution and the dynamics of phytoplankton \cite{mitchell_light_1991, sakshaug_factors_1991, colijn_is_2003}.
So in the following we assume that the nutrient dependence of the growth rate is saturated,
$f_N(N) \rightarrow 1$, and we can neglect the limitation of growth by nutrients.

The spatial profile of light intensity in a water column is described by
Lambert-Beer's law (see e.g. \cite{kirk_light_1994}) which states that the gradient of light intensity at depth $z$ is proportional to the light intensity at this depth
\begin{equation}
\frac{dI}{dz} = -\kappa I \ .
\label{eq:light_decay}
\end{equation}
The coefficient $\kappa$ includes both the absorption of light by water and the attenuation by the phytoplankton cells
$$
\kappa = K_{bg} + k P(z, t) \ ,
$$
where $K_{bg}$ is the background turbidity and $k$ is the {\it per capita} attenuation coefficient of the algae cells. Integrating (\ref{eq:light_decay}) from surface to depth $z$, we obtain
\begin{equation}
I(z) = I_{in} \exp \left[ -K_{bg} z -  \int^z_0  k P(t, z') d z' \right] \ ,
\label{eq:light}
\end{equation}
where $I_{in}$ is the light intensity at surface.


Equations (\ref{eq:phyto}) and (\ref{eq:light}), being coupled by means of the growth rate (\ref{eq:phyto_growth_min}) or (\ref{eq:phyto_growth_mul}), yield an integro-differential system of equations. It is not straightforward to obtain rigorous or analytical results for such a system and even a numerical solution encounters certain difficulties \cite{huisman_population_2002, pham_thi_simulation_2005}.
Nevertheless, without solving any equations, it is clear that the light intensity in the water column is reduced with increasing depth. Thus the light limitation forms a favourable area close to the surface, and the dynamics of the phytoplankton population should be related with the results of the previous chapter, obtained for heterogeneous environments in the presence of advection and diffusion.

\begin{figure}[tb]
   \centerline{\includegraphics[height=6.5cm]{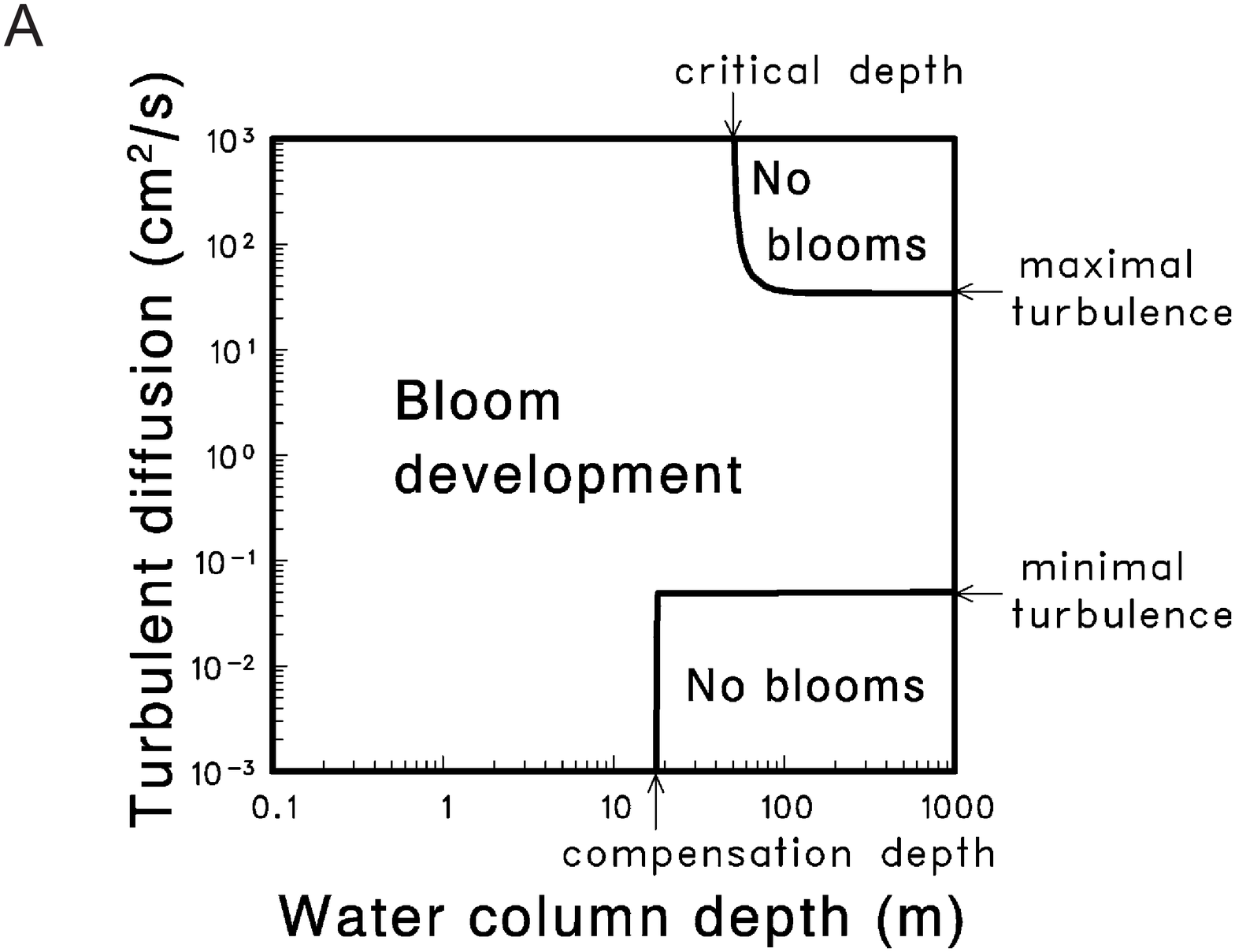}
              \includegraphics[height=6.5cm]{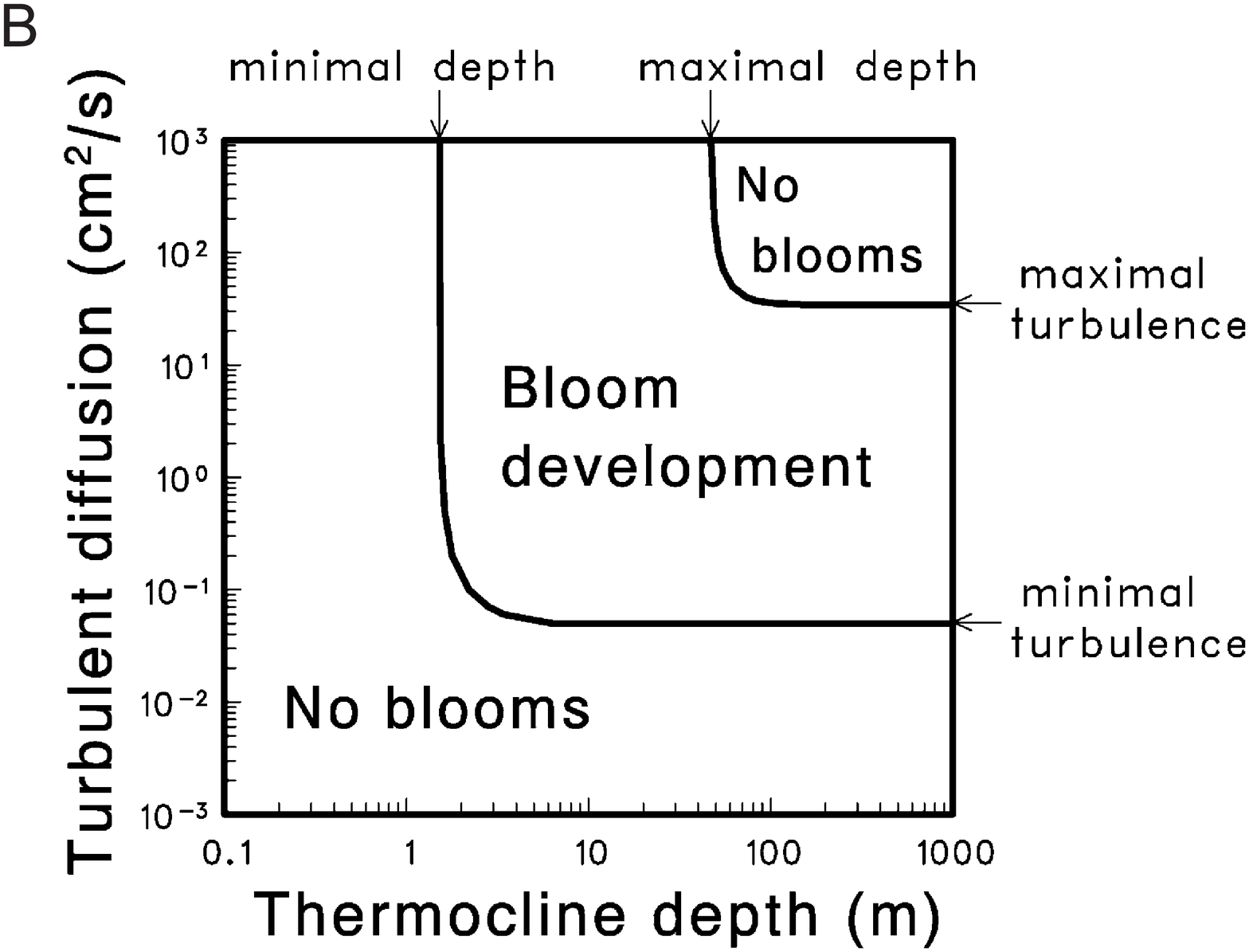}
}
\caption{Critical conditions for phytoplankton blooming: (A) in a closed water column  and (B) in an upper mixed layer. Both figures are reprinted from Journal of Sea Research, {\bf 48}, Huisman and Sommeijer, pp. 83-96 \cite{huisman_population_2002},  Fig.~4 and 6, with permission of the author.
}
\label{fig:cr_depth}
\end{figure}

\paragraph{Critical values for phytoplankton growth}
Depending on the depth of a water column or on the diffusivity, a light limited phytoplankton population can survive or become extinct. Huisman et al. \cite{huisman_do_2002, huisman_population_2002} combined the conditions for survival into a single conception of the critical conditions for phytoplankton blooming in a closed water column (Fig.~\ref{fig:cr_depth}A) and within an UML (Fig.~\ref{fig:cr_depth}B). The main difference between these models is that in a closed system, the sinking of cells is stopped at the bottom, whereas in an UML the biomass can sink across the thermocline to the deep aphotic layers.

\begin{figure}[tb]
\centerline{\includegraphics[height=6cm]{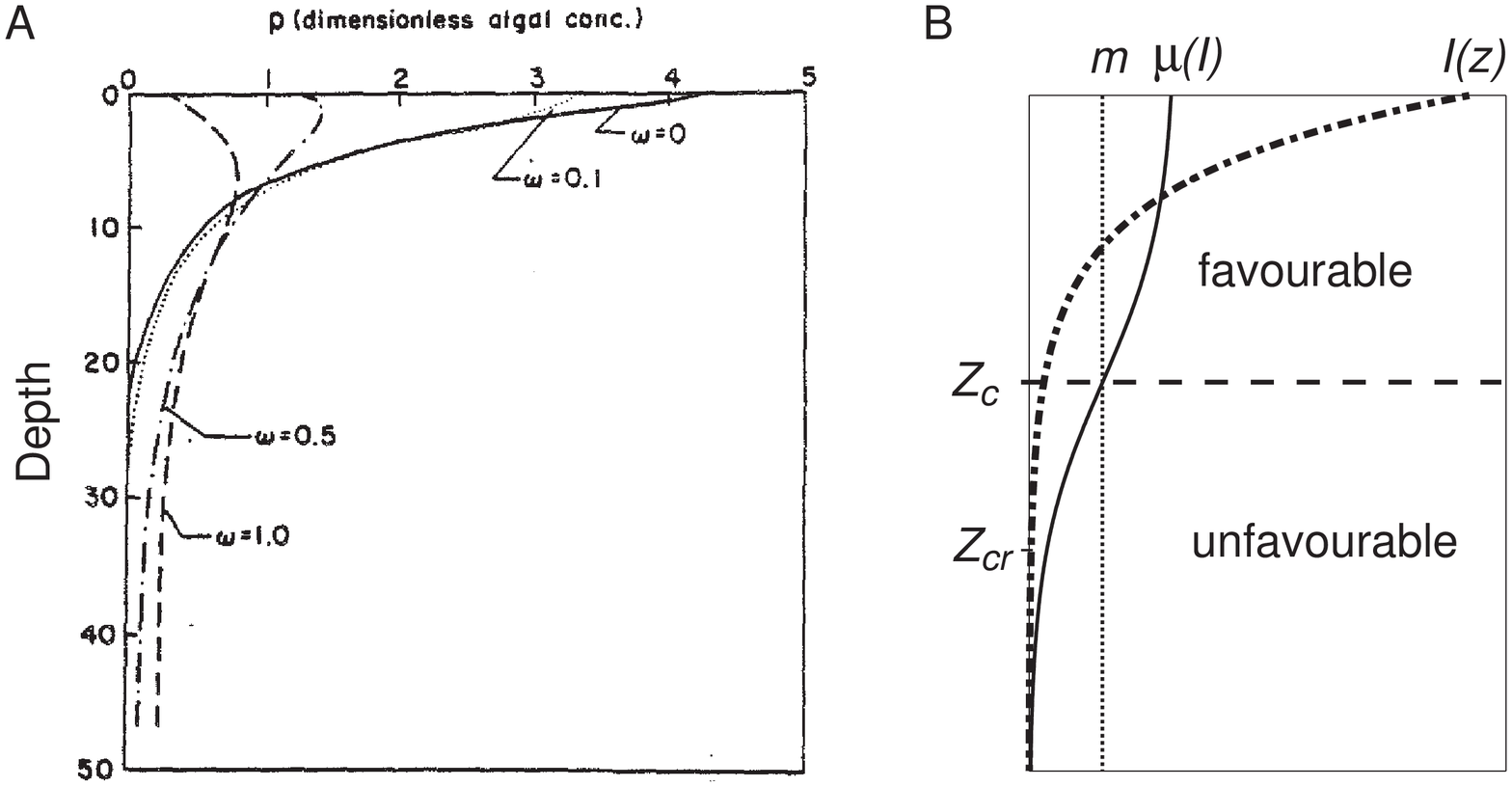}
             }
\caption{(A) Vertical profiles of phytoplankton for different values of the dimensionless sinking velocity $\omega = v/\mu_0 D$ \cite{shigesada_analysis_1981}.
While the shape of the profile is changed, the total amount of biomass in the system remains unchanged. The figure is reprinted from Journal of Mathematical Biology, {\bf 12}, Shigesada and Okubo, 
pp. 311-326 \cite{shigesada_analysis_1981},  Fig.~4,  with permission of the author.
(B) Schematic representation of the compensation depth $Z_c$ and the critical depth $Z_{cr}$.}
\label{fig:LightGrad}
\end{figure}

To determine these conditions, consider first an unstratified water column with a constant diffusivity and assume impenetrable boundary conditions (\ref{eq:boundary_phyto}) for the phytoplankton biomass. Using the system of  equations (\ref{eq:phyto}) and (\ref{eq:light}) in the limit of zero background turbidity, $K_{bg} = 0$, and for a general monotonic growth rate $\mu(I)$, Shigesada and Okubo \cite{shigesada_analysis_1981} showed that a sinking phytoplankton species can establish a population only if
\begin{equation}
D \ge D_{min} = \frac{v^2}{4 (\mu(I_{in}) - m)} \ . \label{eq:phyto_crit_diff}
\end{equation}
This expression coincides with condition (\ref{eq:FK_crit_diff_adv}) from the previous chapter and implies that the {\it minimal diffusivity} should provide a front propagation velocity which is larger than the sinking velocity. The same condition was derived earlier by Riley \cite{riley_quantitative_1949} and other authors from the Fisher-Kolmogorov equation. It is interesting to note \cite{shigesada_analysis_1981} that if $K_{bg} = 0$ and a non-trivial solution exists, then the total biomass in this model does not depend on the sinking velocity. The sinking just shifts the bulk of biomass downward, preserving, however, the total amount of biomass in the water column (Fig.~\ref{fig:LightGrad}A).

Ishii and Takagi \cite{ishii_global_1982} relaxed the condition $K_{bg} = 0$ and proved some existence, stability and uniqueness results for this system. Assuming an algebraic form of the growth rate, $\mu(I) \sim I^\alpha$, Ebert et al. \cite{ebert_critical_2001} have found some  approximation for the minimal diffusivity, $D_{min}$,
and for other critical parameters.

If a water column is sufficiently deep and $K_{bg} \neq 0$ then the net production rate is positive only above the {\it compensation depth}, $Z_c$, which is defined as the depth $z$ at which the local production rate is zero in the absence of biomass (Fig.~\ref{fig:LightGrad}B). From the light attenuation curve (\ref{eq:light}) we find
$$
Z_c = \frac{\ln I_{in} - \ln I_c}{K_{bg}} \ ,
$$
where $I_c$ is defined as the compensation light intensity at which the growth rate is equal to mortality, $\mu (I) = m$, thereby the compensation depth is species specific.

If a water column is shallower than the compensation depth and we assume that the bottom is impenetrable for the biomass, then the population can survive even if mixing is less than a minimal diffusivity (\ref{eq:phyto_crit_diff}) because the cell settling will be stopped at the bottom (Fig.~\ref{fig:cr_depth}A).

In general, the compensation depth divides a water column into a favourable and an unfavourable regime (Fig.~\ref{fig:LightGrad}B). In a well mixed water column the losses in the deep layers can lead to the population extinction. However,
they can be compensated by the production in the euphotic zone, if the unfavourable region is relatively small. Considering a simple mathematical model of a well mixed water column Sverdrup \cite{sverdrup_conditions_1953} defined a {\it critical depth} as the depth of a water column at which the total growth is equal to the total loss of biomass. Similar to the compensation depth, the critical depth can be reinterpreted in terms of the critical light intensity \cite{j._huisman_critical_1999}
$$
Z_{cr} = \frac{\ln I_{in} - \ln I_{out}}{K_{bg}} \ ,
$$
where $I_{out}$ is the light intensity at the bottom of a sufficiently shallow ($Z_B< Z_{cr}$) closed water column after the light limited population of phytoplankton have reached an equilibrium state. The critical depth depends on many parameters, it increases with the incident light intensity and with the phytoplankton growth rate, and it decreases with the mortality rate \cite{j._huisman_critical_1999}.

In a well mixed water column, an excess of the critical depth over the compensation depth determines the maximal possible losses in dark layers, which can be still compensated by the production in the euphotic zone. However, similar to the KiSS model or to the model by Ludwig et al. (see Sec. \ref{sec:crit}) these losses diminish with a decrease of mixing. Extending this research, Huisman et al. \cite{j._huisman_critical_1999} showed that if the depth of a water column or a thermocline exceeds the critical depth, the population survival still is possible if turbulent mixing is less than a {\it maximal diffusivity}. This critical condition is similar to the existence of the maximal diffusivity (\ref{eq:KiSS_Dc}) and (\ref{eq:KiSS_D_Bound}) in the KiSS model, however here it describes the behaviour of a more realistic system. Therefore similar to the KiSS model in advection \eqref{eq:KiSS_D_Bound}, a sinking population can survive in a water column of any depth if mixing remains between a minimal and a maximal value (Fig.~\ref{fig:cr_depth}A).

We now turn to a stratified water column with a UML. We assume that the mixing in the deep layers is less than a critical value \eqref{eq:phyto_crit_diff}, so that the population survival will depend on the characteristics of a UML. Condie and Bormans \cite{condie_influence_1997} showed that if a UML is shallower than
$$
Z_{T, min} \approx \frac{v}{\mu(I_{in}) - m},
$$
a population cannot survive (compare with \eqref{eq:vmax_cond}). In other words, for the survival in a UML, the demographic time scale should be faster than the characteristic time of advection. However usually, $Z_{T, min}$ is  sufficiently small and this criterion is satisfied. In this case the population can persist if the strength of mixing remains within the turbulent window $[D_{min}, D_{max}]$ (see Fig.~\ref{fig:cr_depth}B). Furthermore, if the diffusivity exceeds a maximal value the population survives if the depth of a thermocline is smaller than a {\it maximal depth}, which is defined as the maximal depth of a well mixed upper layer at which losses and production are equal. This depth is slightly smaller than the critical depth in a closed water column, owing to additional losses of biomass across the thermocline.

The fact that a deep upper layer can prevent phytoplankton blooming  was noted experimentally in 1935 by Gran and Braarud \cite{gran_quantitative_1935}, who investigated the conditions of phytoplankton blooming in the upper mixed layer.
They reported that until there exists a deep UML, phytoplankton production cannot exceed the destruction by respiration and phytoplankton blooming is not possible. The concept of the maximal diffusivity is also consistent with field experiments, see e.g. \cite{townsend_spring_1992, ellertsen_spring_1993, peeters_turbulent_2007}.

\subsection{Light and Nutrient limitation}
In the last section we will discuss models which take into account both light and nutrient limitation of phytoplankton growth. These models are more difficult to analyse and often admit only numerical investigation. However, they are more realistic and provide some understanding of the processes occurring in deep waters of many regions where surface layers are nutrient depleted
\cite{radach_vertical_1975, winter_dynamics_1975, platt_modeling_1977, wroblewski_model_1977, walters_time-_1980, varela_simulation_1994, huisman_reduced_2006}. Furthermore, in the tension of two opposing resource gradients the location and the size of a production layer becomes a function of the phytoplankton abundance and the initial conditions, that can lead to new patterns and new dynamical behaviour.

A coupled system of reaction-diffusion equations describing nutrient-phytoplankton cycling was probably first investigated by Okubo \cite{okubo_diffusion-induced_1974, okubo_horizontal_1978}.
Radach and Maier-Reimer \cite{radach_vertical_1975} suggested a mathematical model of phytoplankton growth which included light-nutrient-phytoplankton dynamics. This model was extended by Jamart et al. \cite{jamart_theoretical_1977} who considered limitation by two nutrients, grazing and the variability of the parameters with depth and time. This approach (see also \cite{winter_dynamics_1975, platt_modeling_1977, wroblewski_model_1977, walters_time-_1980}) gave rise to a growing set of ecological models, which include cycling of many chemicals \cite{yakushev_analysis_2007}, coupling with meteorological data \cite{jhnk_summer_2008}, interplay of different phytoplankton groups, and 3D simulations \cite{moll_review_2003, follows_emergent_2007}. Here, however, we will focus on the theoretical aspects and consider only simple conceptual models.

\paragraph{Conservative models}
The nutrient dynamics include uptake by phytoplankton, remineralisation of dead biomass back into a nutrient pool and diffusion. Assuming absolutely effective recycling we obtain
\begin{eqnarray}
\frac{\partial P(z, t) }{\partial t} &=& \mu(N, I)  P - m P - v \frac{\partial P }{\partial z} +
D \frac{\partial^2 P  }{\partial z^2} \ , \nonumber \\ \label{eq:phytoHR}\\
\frac{\partial N(z, t) }{\partial t} &=& -\mu(N, I)  P + m P  + D \frac{\partial^2 N  }{\partial z^2} \ , \nonumber
\end{eqnarray}
where the biomass is measured in terms of its nutrient content (compare to the non-spatial version (\ref{eq:NP}) of this model). We do not include advection in the second equation, as nutrients, which are dissolved in water, are only slightly influenced by the gravity force. Nevertheless, this term should appear, if advection is caused by a vertical or horizontal stream.

Furthermore, we assume that the nutrient cannot diffuse across the surface and a large nutrient pool in the sediment or in deep ocean layers sustains a constant concentration, $N_B$, the bottom of the water column
\begin{equation}
   \frac{\partial N(0, t)}{\partial z} = 0 \ ,
   \qquad N(Z_B, t) = N_B \ .
\label{eq:boundary_nutr}
\end{equation}

\begin{figure}[bt]
  \centerline{\includegraphics[height=5.5cm]{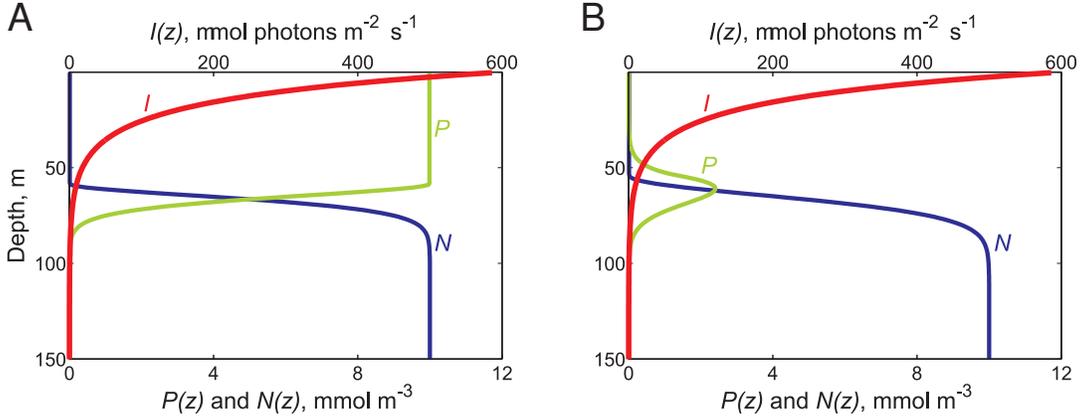}}
\caption{(Colour online) Typical distributions of phytoplankton (green), nutrient (blue), and light (red) in the conservative model (\ref{eq:phytoHR}) without sinking (A) and with sinking (B), according to Hodges and Rudnick \cite{hodges_simple_2004}.}
\label{fig:hodges}
\end{figure}

Fig.~\ref{fig:hodges} shows typical final distribution of phytoplankton and nutrient given by model \eqref{eq:phytoHR}, supplemented by equation \eqref{eq:light} for light. Hodges and Rudnick \cite{hodges_simple_2004}
pointed out that, independent of the functional form of the growth rate and of the light distribution (assuming that light decreases with depth), this model can reproduce a deep stationary phytoplankton maximum only if $v>0$. In other words, the presence of opposing resource gradients is not sufficient to reproduce a deep phytoplankton maximum. To prove this, let us define the total concentration of the nutrient as $S = P + N$. Consider an equilibrium state, when the left-hand-side of (\ref{eq:phytoHR}) equals zero. By adding both equations (\ref{eq:phytoHR}) we obtain
$$
D \frac{\partial^2 S  }{\partial z^2} - v \frac{\partial P }{\partial z} = 0 \ .
$$
Assuming $v = 0$ and integrating this equation over $z$ we find
$$\frac{\partial S  }{\partial z} = {\rm const} = \frac{\partial  P }{\partial z} + \frac{\partial N  }{\partial z} = 0 \ ,
$$
owing to the boundary condition (\ref{eq:boundary_phyto}) and (\ref{eq:boundary_nutr}) at the surface. Thus $S = N + P = {\rm const}$ and a deep phytoplankton maximum should be accompanied by a deep minimum of nutrient. However, if the light intensity reduces with depth, this profile is unstable because there is no factor limiting phytoplankton growth in the upper layer. Thus this system should exhibit a surface maximum (Fig.~\ref{fig:hodges}A). However, similar to the model without nutrient limitation (Fig.~\ref{fig:LightGrad}A), the phytoplankton sinking shifts the maximum of biomass downwards (Fig.~\ref{fig:hodges}B).

Extending this model, Hodges and Rudnick \cite{hodges_simple_2004} included a detrital pool, $T(z, t)$, as the third compartment
\begin{equation}
\begin{array}{lrr}
\displaystyle\frac{\partial P(z, t) }{\partial t} =
& \mu(N, I)  P - m P
&- v_P \displaystyle\frac{\partial P }{\partial z}
+ D \displaystyle\frac{\partial^2 P  }{\partial z^2} \ ,  \\ \\
\displaystyle\frac{\partial T(z, t) }{\partial t} =&
 m P - r T &
 - v_T \displaystyle\frac{\partial T }{\partial z}
+D \displaystyle\frac{\partial^2 T  }{\partial z^2} \ , \\  \\
\displaystyle\frac{\partial N(z, t) }{\partial t} =&
 -\mu(N, I)  P + r T  &
 + D \displaystyle\frac{\partial^2 N  }{\partial z^2} \ ,
\end{array}
\label{eq:three_cmpt}
\end{equation}
where $v_P$ and $v_T$ are the sinking velocities of phytoplankton and detritus respectively. Note that usually detritus sinks much faster than phytoplankton \cite{shanks_marine_1980, helle_ploug_ballast_2008}.
In this model the cycle of chemicals includes three stages: the transfer of biomass to detritus with mortality $m$, the remineralisation of detritus back into nutrients with remineralisation rate $r$, and finally the consumption of nutrient by biomass. While this model can exhibit deep maxima if $v = 0$, the change of phytoplankton concentration is very small and cannot represent real data. An apparent maximum can be observed only if one assumes sinking of detritus or phytoplankton. Hodges and Rudnick extended this statement to any number of compartments, which however do not include depth dependent parameters. Thus, sinking is a major component of this system. The sedimentation of organic matter removes  the nutrient fixed in phytoplankton cells from the upper layer, which  leads to the formation of deep phytoplankton maxima.

\begin{figure}[tb]
   \centerline{\includegraphics[height=8cm]{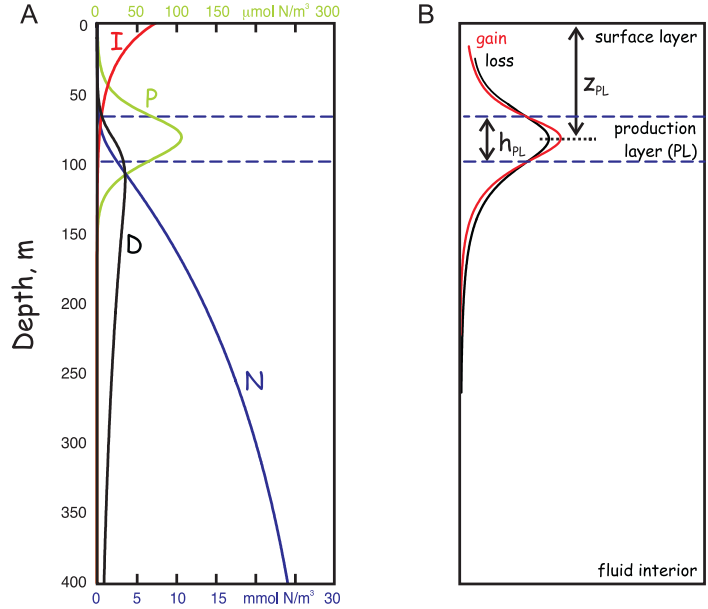}}
\caption{(Colour online) Typical distributions in the three compartments model \eqref{eq:three_cmpt}. (A) Vertical profiles of phytoplankton ($P$), detritus ($D$), nutrient ($N$), and light ($I$).
(B) The upper and lower limits of the production layer (dashed lines) are  the species compensation depths. The figures are reprinted from Progress in Oceanography, {\bf 75}, Beckmann and Hense, pp. 771-796 \cite{beckmann_beneath_2007},  Fig.~2, with permission of the author.}
\label{fig:beckman}
\end{figure}

Beckmann and Hense \cite{beckmann_beneath_2007} performed numerical simulations and analytical evaluations of model \eqref{eq:three_cmpt}, assuming that detritus sinks relatively fast, whereas the phytoplankton sinking is negligible. Fig.~\ref{fig:beckman} reproduces a typical distribution of physical characteristics in this model. Furthermore, Beckmann and Hense suggested to extend the concept of compensation depths. Instead of the static definition in the absence of biomass they suggested to use two dynamical depths at which the {\it in situ} production rate of phytoplankton is zero, owing to the light or nutrient limitation. If phytoplankton sinking velocity is zero then in equilibrium  (see \eqref{eq:three_cmpt}) these values can be expressed from the phytoplankton distribution
$$
 \left. \frac{\partial^2 P}{\partial z^2} \right|_{Z^{(N)}_c} = 0 \ ,
 \qquad
  \left. \frac{\partial^2 P}{\partial z^2} \right|_{Z^{(I)}_c} = 0 \ ,
$$
where $Z^{(N)}_c$ and $Z^{(I)}_c$ are the compensation depths due to nutrient and light limitation (Fig.~\ref{fig:beckman}B).

\paragraph{Non-conservative models}
Model \eqref{eq:three_cmpt} contains three reaction-diffusion equations and an equation for the light distribution. This makes further analysis difficult. However, since detritus sinks relatively fast \cite{shanks_marine_1980, helle_ploug_ballast_2008}, we can simplify the model assuming that a part $\varepsilon$ of the dead biomass is instantly remineralised {\it in situ}, whereas the rest sinks until it reaches the bottom and sustains a constant nutrient concentration at the bottom. Thus we obtain the following non-conservative system of equations
\begin{eqnarray}
\frac{\partial P(z, t) }{\partial t} &=& \mu(N, I)  P - m P - v \frac{\partial P }{\partial z} +
D \frac{\partial^2 P  }{\partial z^2} \ , \nonumber \\
\label{eq:phytoHREPS}\\
\frac{\partial N(z, t) }{\partial t} &=& - \mu(N, I)  P + \varepsilon m P  + D \frac{\partial^2 N  }{\partial z^2} \ . \nonumber
\end{eqnarray}
This model (compare to equation (\ref{eq:NP_noncons})) can reproduce
deep phytoplankton maxima even if the sinking velocity is zero, owing to the fact that a part $(1 - \varepsilon)$ of the fixed nutrient  is implicitly transferred from the upper layer to the bottom. Even though this model is non-conservative and has apparent disadvantages, it or similar models were successively applied to reproduce field data  \cite{jamart_theoretical_1977, varela_simulation_1994,  huisman_reduced_2006, yoshiyama_catastrophic_2002}. However, we are not aware of any comparison of the two model classes \eqref{eq:phytoHR} and \eqref{eq:phytoHREPS}, which might be interesting.

In the case of zero sinking velocity, Klausmeier and Litchman \cite{klausmeier_algal_2001} performed analytical calculations for model \eqref{eq:phytoHREPS}. Assuming that the phytoplankton distribution can be approximated by a Dirac $\delta$-function and further that an infinitely small production layer should be located to balance the light and the nutrient limitation, Klausmeier and Litchman found an equation for the position $Z^*$ of a deep maximum, which for boundary conditions \eqref{eq:boundary_phyto} and \eqref{eq:boundary_nutr} reads as
$$
\frac{\ln \left(I_{in}/{I_c}\right)}{k} - \frac{K_{bg}}{k} Z^* = \frac{\mu_0 D (N_B - N_c)}{m (1 - \varepsilon) (Z_B - Z^*)} \ ,
$$
where $N_c$ and $I_c$ are the critical values of light and nutrient intensity for which the growth rate is equal to the mortality rate.

\begin{figure}[tb]
\centerline{
\includegraphics[height=10cm]{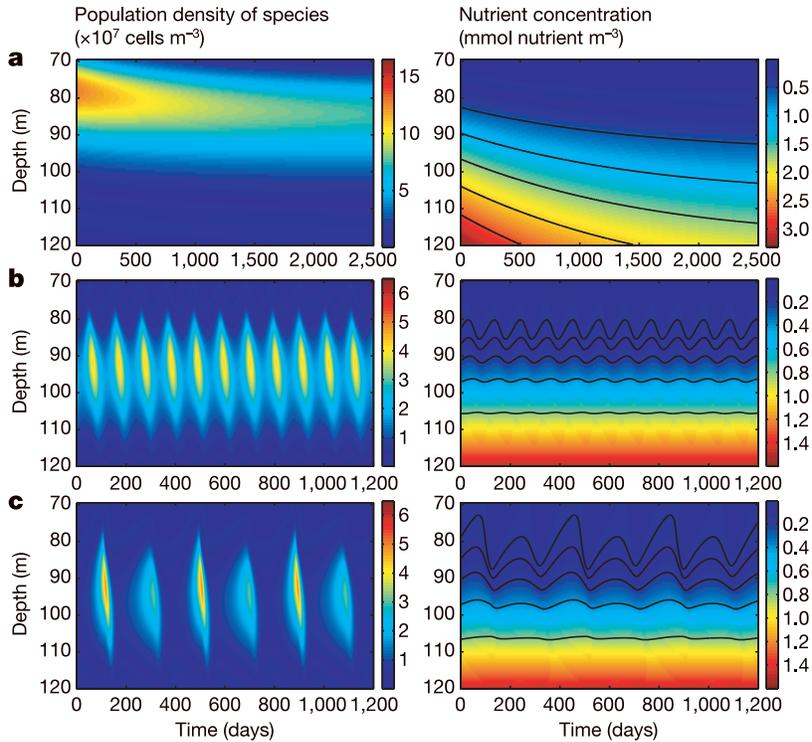}}
\caption{(Colour on line) Evolution of the phytoplankton density and the nutrient concentration with time. (a) A stable DCM ($D = 0.5 \ {\rm cm^2/s}$), (b) small oscillations in the DCM ($D = 0.2 \ {\rm cm^2/s}$), and (c) large-amplitude oscillations in the DCM, with double periodicity ($D = 0.12 \ {\rm cm^2/s}$). The figures are reprinted from Nature, {\bf 439}, Huisman et al., pp. 322-325 \cite{huisman_reduced_2006},  Fig.~2, with permission of the author.
}
\label{fig:huisman}
\end{figure}

\paragraph{Oscillations and chaos}
Huisman et al. \cite{huisman_reduced_2006} pointed out that system \eqref{eq:phytoHREPS} exhibits oscillations of biomass if the mixing is reduced below a critical value. Fig.~\ref{fig:huisman} shows the  behaviour of biomass and of nutrient in two typical cases.
In the first case (Fig.~\ref{fig:huisman}a) the mixing intensity is high enough to provide a stable distribution of biomass. If however the level of diffusivity is reduced, then only oscillatory, or even chaotic patterns, can appear (see Fig.~\ref{fig:huisman}b and \ref{fig:huisman}c). As noted by Huisman et al. these oscillations are caused by the difference in the time scales of the rapid transport of phytoplankton, consuming the nutrients, and the slow upward transport of nutrients. Furthermore, as shown in Section~\ref{Seq:DriftPar}, for the survival of a population in an advective flux the diffusivity should exceed a minimal level (\ref{eq:KiSS_D_Bound}),
which increases with the reduction of the habitat and of the growth rate.
In the absence of biomass the nutrient can be nearly uniformly distributed over the water column, thereby the growth rate becomes only light limited and the production layer extends from the surface to the compensation depth, which is usually sufficiently large. Thus, without biomass, the level of mixing might be sufficient to induce population growth. However, the consumption of nutrients and self-shading of light reduce both the growth rate and the width of the production layer. That, in turn, increases the value of the minimal diffusivity (\ref{eq:KiSS_D_Bound}) and finally the sinking may lead to the population wash-out if the diffusivity in the water column becomes insufficient.

Koszalka et al. \cite{koszalka_plankton_2007} noted that the periodic oscillations of phytoplankton biomass will be most probably disguised by currents and horizontal inhomogeneity in a real ecosystem.

\paragraph{Upper mixed layer}
Hodges and Rudnik \cite{hodges_simple_2004} and Beckmann and Hense \cite{beckmann_beneath_2007} showed that if self-shading of light can be neglected in equation (\ref{eq:light}), then an upper mixed layer does not lead to any qualitative changes in the system dynamics. However, Yoshiyama and Nakajima \cite{yoshiyama_catastrophic_2002} pointed out that a UML can lead to bistability of phytoplankton profiles.

\begin{figure}[tb]
   \centerline{\includegraphics[height=8cm]{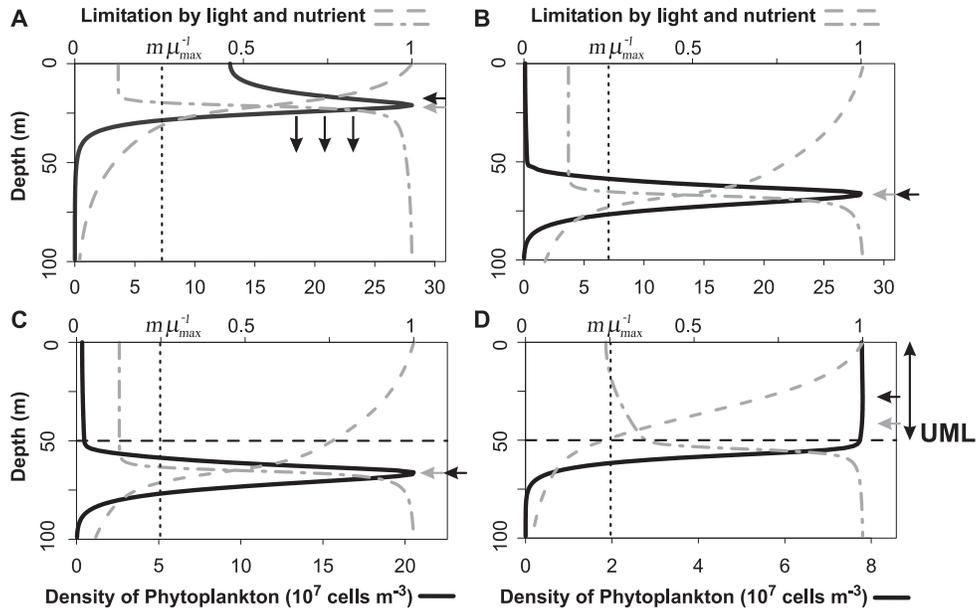}
}
\caption{Typical vertical phytoplankton profiles $P(z)$ in a system without a UML (top) and with a UML (bottom), assuming a gradual change of diffusivity \eqref{eq:Dz} in model \eqref{eq:phytoHREPS}. Without a UML, a non-stable phytoplankton distribution (A) evolves to a single stable solution (B). Under the same conditions in the system with a UML, we observe two stable distributions:
with a maximum in the deep layers (C) and with a maximum in the UML (D).
The dot-dashed and dashed lines show the limitations of growth \eqref{eq:Menten} by light and by nutrient, respectively, vertical dotted line shows the level of mortality.
Black and grey arrows show the centers of biomass and net production, respectively.}
\label{fig:centers}
\end{figure}

Ryabov et al. \cite{ryabov_bistability} generalised this result by taking into account the competition of two species and relaxing other assumptions. They considered the model \eqref{eq:phytoHREPS}, assuming a gradual change of diffusivity \eqref{eq:Dz} from a UML to the deep layers, and showed that under certain parameters, depending on the initial conditions the production layer can be steadily located either within a UML or below it. Fig.~\ref{fig:centers} provides a rough insight into the system dynamics. In the absence of an upper mixed layer the difference in the locations of biomass and of the production layer drives the bulk of biomass towards the production layer (Fig.~\ref{fig:centers}A). The shift of biomass can lead to the redistribution of resources, which in turn can change the location of the production layer. This process repeats until the system reaches an equilibrium configuration (Fig.~\ref{fig:centers}B), when the centre of biomass coincides with the centre of production. Now consider a system with an UML. In a certain range of parameter the UML does not affect distributions with a deep maximum of biomass (Fig.~\ref{fig:centers}C). However, the initial growth of biomass within the UML begets another stable solution with a maximum of biomass located within the UML. The biomass is almost uniformly distributed within the UML and its location is uncoupled from the location of the production layer (Fig.~\ref{fig:centers}D). As a result, a gradual shift of the bulk of biomass into deep layers is no longer possible and the transition to a deep biomass maximum can only take place if the light intensity below the UML is sufficiently large to provide positive net growth in deep layers -- otherwise the phytoplankton remains trapped in the UML. Thus the production layer can occupy different parts of the water column, depending on the current system state and on initial conditions.

\section{Discussion}

Concluding this review we would like to make a few notes. First, let us  compare the behaviour of the critical patch models and those based on the consumer-resource dynamics. The latter models can be divided into two large groups. In the first group we would include those systems in which the location of a favourable patch is constrained by some environmental conditions. For instance, the limitation of phytoplankton growth by light leads to the formation of the favourable patch in the upper level of a water column.
The dynamics of this group and of the critical patch models demonstrate many general traits
and many effects can by predicted and evaluated on the basis of the minimal models. The second group consists of the models in which the location of the favourable patch is determined by the dynamical interplay of different factors. For example, we can consider the growth of phytoplankton biomass driven by two opposing resource gradients. In this group, the location of the favourable patch is not predefined. Moreover, the system dynamics becomes very sensitive to the implementation of the consumer-resource cycling. This complexity leads to the arising of new patterns and new dynamical behaviour, which can  hardly be reproduced in the framework of the critical patch models.

The second remark concerns the advantages and disadvantages of partial differential equations (PDEs) for modelling ecological systems.
PDEs provide very convenient and powerful tools for the investigation of population dynamics. First, in the same framework, we can consider such different and complex phenomena as, for instance, the vertical distribution of sinking phytoplankton cells or the survival of a population drifting in a flow. 
Second, analytical solutions in many cases provide important predictions and understanding of the main effects, which can appear in more realistic systems. Third, one can perform an exhaustive numerical simulation of a model, determining all possible bifurcation points.
Finally, seemingly the pool of methods developed for the analysis of partial differential equations is not played out yet and this approach can still gain a lot of useful techniques from quantum mechanics and statistical physics.
However, we would like to mention as well some restrictions of this approach.  Intrinsically it is always suggested that this approach
is suitable for systems containing many organisms, so that the relative fluctuations of density become negligible and all function are continuous. However this statement does not hold if we consider the survival-extinction transition. As the system approaches its critical state, the population density declines and the fluctuations of density (demographic stochasticity) start to play a crucial role \cite{tilman_spatial_1997}. Thus, in reality, the extinction of a population might occur under conditions which still allow for the population survival in a deterministic PDE framework. Therefore, the development of a theory including stochastic effects is necessary for the correct representation of the transient behaviour.

\section*{Acknowledgements}
We are grateful to Aike Beckmann, Jef Huisman, and two anonymous referees for their helpful comments on the manuscript.


\end{document}